\RequirePackage{fixltx2e}
\documentclass[prd,twocolumn,showpacs,preprintnumbers,superscriptaddress,nofootinbib,amsmath,amssymb,floatfix,nobalancelastpage]{revtex4}
\usepackage{graphicx}
\usepackage{bm}
\usepackage{dcolumn}
\usepackage{amsmath}
\usepackage{amssymb}
\usepackage{color}
\usepackage{url}
\usepackage{makecell}

\usepackage{aas_macros}

\usepackage{hyperref}

\usepackage{subfigure}

\def\apjl{ApJL}
\def\apjs{ApJS}

\def\aap{A \& A}

\def\arccosh{{\rm Arccosh}}

\def\arcsinh{{\rm Arcsinh}}
\def\arctanh{{\rm Arctanh}}
\def\arctan{{\rm Arctan}}
\def\arcsin{{\rm Arcsin}}

\def\rh{R_{\rm h}}
\def\mh{m_{\rm h}}
\def\Mh{M_{\rm h}}
\def\sigmalos{\sigma_{\rm los}}
\def\rd{R_{\rm d}}
\def\rs{r_{\rm s}}
\def\rt{r_{\rm t}}

\def\kpc{\,{\rm kpc}}

\def\kms{\,{\rm km\,s^{-1}}}
\def\dex{\,{\rm dex}}

\def\percent{\text{ per cent}}

\def\rhoDM{{\rho_{\rm DM}}}

\def\Rd{{R_{\rm d}}}
\def\Rc{{R_{\rm c}}}
\def\rhoDM{{\rho_{\rm DM}}}

\def\sigv{{\langle\sigma v\rangle}}

\def\percent{\text{ per cent}}

\newcommand{\ds}{{\displaystyle}}

\newcommand{\Df}{{\rm D}}
\newcommand{\Jf}{{\rm J}}

\begin{document}

\title{Indirect Dark Matter Detection for Flattened Dwarf Galaxies}

\author{Jason~L. Sanders}
\email{jls@ast.cam.ac.uk}
\affiliation{Institute of Astronomy, Madingley Rd, Cambridge, CB3 0HA}

\author{N.~Wyn Evans}
\email{nwe@ast.cam.ac.uk}
\affiliation{Institute of Astronomy, Madingley Rd, Cambridge, CB3 0HA}

\author{Alex Geringer-Sameth}
\email{alexgs@cmu.edu}
\affiliation{McWilliams Center for Cosmology, Department of Physics, Carnegie Mellon University, Pittsburgh, PA 15213}

\author{Walter Dehnen}
\email{wd11@leicester.ac.uk}
\affiliation{Department of Physics \& Astronomy, University of Leicester, University Road, Leicester, LE1 7RH}

\date{\today}

\begin{abstract}
Gamma-ray experiments seeking to detect evidence of dark matter annihilation in dwarf spheroidal galaxies require knowledge of the distribution of dark matter within these systems. We analyze the effects of flattening on the annihilation (J) and decay (D) factors of dwarf spheroidal galaxies with both analytic
and numerical methods. Flattening has two consequences: first, there
is a geometric effect as the squeezing (or stretching) of the dark
matter distribution enhances (or diminishes) the J-factor; second, the
line of sight velocity dispersion of stars must hold up the flattened baryonic
component in the flattened dark matter halo. We provide analytic formulae and a simple numerical approach to estimate the correction to the J- and D-factors required over simple spherical modeling. The formulae are validated with a series of equilibrium models of flattened stellar distributions embedded in flattened dark-matter distributions.
We compute corrections to the J- and D-factors for the Milky Way dwarf spheroidal galaxies under the assumption that they are all prolate or all oblate and find that the hierarchy of J-factors for the dwarf spheroidals is slightly altered (typical correction factors for an ellipticity of $0.4$ are $0.75$ for the oblate case and $1.6$ for the prolate case). We demonstrate that spherical estimates of the D-factors are very insensitive to the flattening and introduce uncertainties significantly less than the uncertainties in the D-factors from the other observables for all the dwarf spheroidals (for example, ${}^{+10\percent}_{-3\percent}$ for a typical ellipticity of $0.4$). We conclude by investigating the spread in correction factors produced by triaxial figures and provide uncertainties in the J-factors for the dwarf spheroidals using different physically-motivated assumptions for their intrinsic shape and axis alignments. We find that the uncertainty in the J-factors due to triaxiality increases with the observed ellipticity and, in general, introduces uncertainties of a factor of $2$ in the J-factors. We discuss our results in light of the reported gamma-ray signal from the highly-flattened ultrafaint Reticulum II. Tables of the J- and D-factors for the Milky Way dwarf spheroidal galaxies are provided (assuming an oblate or prolate structure) along with a table of the uncertainty on these factors arising from the unknown triaxiality.
\end{abstract}


\maketitle

\section{Introduction}

In recent years, gamma-ray observations of Milky Way dwarf spheroidal galaxies (dSphs) have
led to great strides in sensitivity to dark matter annihilation. Here
the goal is to probe particles which interact with the Standard Model with the well-motivated weak-scale annihilation
cross section $\sigv \simeq 3 \times 10^{-26}\,\mathrm{cm^3 \,s^{-1}}$. Particles having this cross section will exist today with an abundance equal to that observed for dark matter $\Omega_{\rm DM}$, making this so-called relic cross section a natural target for experimental searches for annihilation.
Combined analyses of dSphs using data from
Fermi Large Area Telescope (LAT) first ruled out the relic cross section for dark matter particle masses of a few tens of
GeV~\cite{2011PhRvL.107x1303G,2011PhRvL.107x1302A} and follow-up
analyses incorporating more dSphs and increased observation time
continue to improve
sensitivity~\cite[e.g.][]{2014PhRvD..89d2001A,2015PhRvD..91h3535G,2015PhRvL.115w1301A}. For
higher dark matter masses ($M \gtrsim \mathrm{TeV}$), the three major
Cherenkov telescope collaborations continue to invest significant time
on pointed observations of Milky Way dSphs. The resulting upper limits
are two to three orders of magnitude from the relic cross
section~\cite{2012PhRvD..85f2001A,2014JCAP...02..008A,2014PhRvD..90k2012A,2016JCAP...02..039M,2015arXiv150901105Z},
but the situation bodes well for the future CTA
project~\cite[e.g.][]{2015arXiv150806128C}.

An exciting development in this field is the recent and
ongoing discovery of large numbers of new Milky Way satellites made
possible by wide-area photometric
surveys~\cite[e.g.][]{2015ApJ...805..130K,2015ApJ...807...50B,2015ApJ...802L..18L,2015ApJ...813..109D}. Since
2015 the number of known Milky Way satellites has approximately
doubled thanks to Southern hemisphere data from the Dark Energy Survey
and Pan-STARRS. These new dSphs have the potential to significantly
build on current efforts to uncover evidence of dark matter
annihilation~\cite[e.g.][]{2015PhRvD..91f3515H,2015PhRvL.115h1101G,2015ApJ...809L...4D,2015JCAP...09..016H,2015arXiv151109252L}.

Intriguingly, the first of these new dwarf spheroidal galaxies discovered, Reticulum~II,
shows indications of a gamma-ray signal exceeding background in the
Fermi-LAT data~\cite{2015PhRvL.115h1101G}.  Two methods of modeling
the gamma-ray background yield false-alarm probabilities of $p=0.0001$
and $0.01$ for detecting such a signal. Subsequent
analysis~\cite{2015JCAP...09..016H} confirmed the results
of~\cite{2015PhRvL.115h1101G} and argued that the Reticulum~II signal
was consistent with the gamma-ray excess reported from the Galactic
Center and claimed as dark matter. With a reprocessing of the raw
Fermi data~\cite{2015ApJ...809L...4D}, the Fermi-LAT Collaboration
found an increased probability for a background fluctuation explaining
the Reticulum~II signal ($p=0.05$) and concluded the signal is
insignificant. Making sense of the results
of~\cite{2015PhRvL.115h1101G} and~\cite{2015ApJ...809L...4D} is
complicated by the fact that the two datasets are only partially
independent, sharing approximately half the detector events. A
separate analysis is needed to compute joint probabilities of
background fluctuation in the partially correlated datasets.

In this work we follow a different path towards assessing dark matter interpretations of gamma-ray signals.
Rather than analyzing the gamma-ray data, we consider the determination of the dark matter
content of the Milky Way's dSphs, a necessary ingredient for performing optimized combined searches using
dSphs. A critical test of any alleged dark
matter signal from dSphs is that the amplitude of the gamma-ray signal must
scale amongst the dSphs according to their J-factors (see,
e.g.,~\cite{2015ApJ...801...74G,2015MNRAS.453..849B}). The J-factor is
the square of the dark matter density integrated along the line of
sight and over the solid angle of the observation,
\begin{equation}
J = \int\int \rhoDM^2(\ell, \Omega) d\ell d\Omega.
\label{eq:Jfactor}
\end{equation}
While annihilating dark matter models are theoretically better motivated, there are models in which dark matter decays \citep{Ibarra2013}. In these models, the relevant astrophysical factor is the D-factor, which is the dark matter density integrated along the line of
sight and over the solid angle of the observation.

Robust determinations of the relative J-factors is of prime
importance. For instance, the Fermi-LAT
Collaboration~\cite{2015ApJ...809L...4D}, under the assumption that
each of eight considered dSphs was equally likely to produce a signal,
further diluted the significance of the Reticulum~II gamma-ray excess to $p=1 - (1-0.05)^8 = 0.33$, concluding that it is
insignificant. However, there are reasons to doubt the usefulness of
this argument as Reticulum~II is closer and very highly flattened,
both of which can enhance the amplitude of an annihilation signal compared
to other dSphs. Therefore, we require accurate relative estimates of the J- and D-factors, but unfortunately the
data on the most tempting dSph candidates are often of limited quality. Motivated
by this \cite[][~hereafter Paper I]{EvansSandersGS} provided simple formulae for
the J- and D-factors for a spherical NFW profile and infinite
spherical cusps. The formulae relied on the empirical law that the
mass within the half-light radius is well constrained as \citep{Wa09,Wo10}
\begin{equation}
\Mh=M(\rh)\approx \frac{5}{2G}\langle\sigma_{\rm los}^2\rangle \rh,
\label{eq:wolf}
\end{equation}
where $\rh$ is the (projected) half-light radius of the stars and
$\langle\sigma_{\rm los}^2\rangle$ is the luminosity weighted squared line-of-sight velocity dispersion.

However, an entirely characteristic feature of dSphs is in the name --
spheroidal! They are flattened (with a typical ellipticity between $0.3$ and $0.5$), and some of the ultrafaints are very
highly flattened with ellipticities exceeding $0.5$, such as
Hercules~\citep{De12}, Ursa Major I~\citep{Ma08}, Ursa Major
II~\citep{Zu06}, and indeed Reticulum~II~\citep{Ko15}. Therefore, the
underlying physical model of a spherical dark halo containing a round
distribution of stars may fail to capture important aspects of the
physics. Here we extend the scope of spherical analyses, to account
for the effects of flattening in both the stellar and dark matter
profiles. \citet{Bo15d} provided a systematic investigation of J-factors of flattened figures. Here, two mildly triaxial numerical
models of dSphs (created for The {\it Gaia} Challenge) were viewed
along each of the short, medium and long axes. This investigation
revealed that the projection effects can have a significant impact on
the velocity dispersion, and concluded that the J-factors constructed
by Jeans analyses can vary from the true values by $\sim 2.5$. Recently, \cite{Hayashi2016} computed J-factor estimates for the dSphs using axisymmetric Jeans modelling. These authors attributed the differences between their measured J-factors and those from spherical analyses primarily to other modelling assumptions.

It is natural to expect that the dissipationless
dark matter distribution is rounder -- or at least no more flattened
-- than the dissipative baryonic component. So, large classical dSphs
which appear roundish on the sky (such as Leo I and II) may have
almost spherical dark matter halos. However, the dark halos of the ultrafaints are expected to be more
highly flattened than those of the classical dSphs, as it is known
that baryonic feedback effects drive the dark matter distribution
towards sphericity~\citep{Ab10,Ze12}. The ultrafaints have such a puny
baryonic content that pure dissipationless
simulations~\citep{Ji02,Al06}, which find strongly triaxial and nearly
prolate dark halos, may be a much better guide to the true shape. For instance, recent simulations have found that
the baryonic distribution is just $\sim10\percent$ flatter than the
dark-matter distribution for dark-matter halos of $10^{10} M_\odot$
~\citep{Tenneti2014}. Throughout this paper, we work under the assumption that the dark matter distribution is flattened in the same way as the stellar
distribution.

The effects of flattening can be understood qualitatively for a few
simple configurations. The simplest is the face-on case when the
dark-matter and stellar distributions are flattened along the
line-of-sight. Observationally, the isophotes still appear circular
and the measured half-light radius remains the same, but we have
increased (decreased) the density of dark matter in the oblate
(prolate) case. Naturally, this effect -- which we refer to as the
\emph{geometric factor} -- gives rise to a larger (smaller) J-factor
than a spherical analysis would infer. But, we must also consider the
effect of flattening on the line-of-sight velocity dispersion, which
we call the \emph{kinematic factor}. For the oblate case, the stellar
distribution is more compressed, so the line-of-sight dispersion is
now smaller than the spherically averaged dispersion. Less contained
mass is inferred and so the spherical J-factor underestimates the
total J-factor. Therefore, for face-on viewing of an oblate figure,
both the geometric and kinematic effects cause the J-factor inferred
from a spherical analysis to be an underestimate of the true value.
For the prolate case, the velocity dispersion is larger than the
spherically-averaged dispersion and so more mass is inferred and the
spherical J-factor is an overestimate.

When the dSph is viewed \emph{edge-on} such that it appears flattened
in the sky, the combined result of the kinematic and geometric effects
is less clear. For oblate figures, the density is increased over the
spherical case, whilst the half-light radius remains the same. These
geometric effects cause the J-factor assuming sphericity to be an
underestimate.  However, the kinematic factor works the other way, as
the measured velocity dispersion is greater than the spherical
average. We will see that the combination of these two competing
effects leads to a small decrease in the true J-factor over that
inferred from a spherical analysis. For the prolate case, we have the
converse situation with the geometric factor leading to an
overestimate whilst the kinematic factor leads to an
underestimate. However, now the stretching of the stellar profile in
the sky causes the half-light radius to increase. We will see that the
net result is an increase in the true J-factor over the spherical
J-factor.

This qualitative explanation is tested in Section II where we
construct equilibrium models of the Reticulum~II galaxy via the
made-to-measure method. We explore a range of different flattenings
and provide simple fits for the correction factors. In Section III, we
use these fits to derive J-factors for the known dwarf spheroidals under
the assumption that they are either prolate or oblate. In Section IV,
we build intuition for our numerical results by considering two
families of axisymmetric equilibria for which analytic progress is
possible and present a more rapid general approach for estimating the correction factors using the virial theorem. Section V extends these findings to the triaxial case and
demonstrates how the correction factors vary for a triaxial figure as a
function of the viewing angle. In Section VI, we discuss the
constraints and evidence on the intrinsic shapes and alignments of the
Milky Way dSphs and give estimates of the uncertainties in the J-factors of the dSphs due to unknown triaxiality. In Section VII we summarize our findings and discuss possible implications for the claimed signal from Reticulum~II in light of our work.

\begin{figure}
\includegraphics[width=\columnwidth]{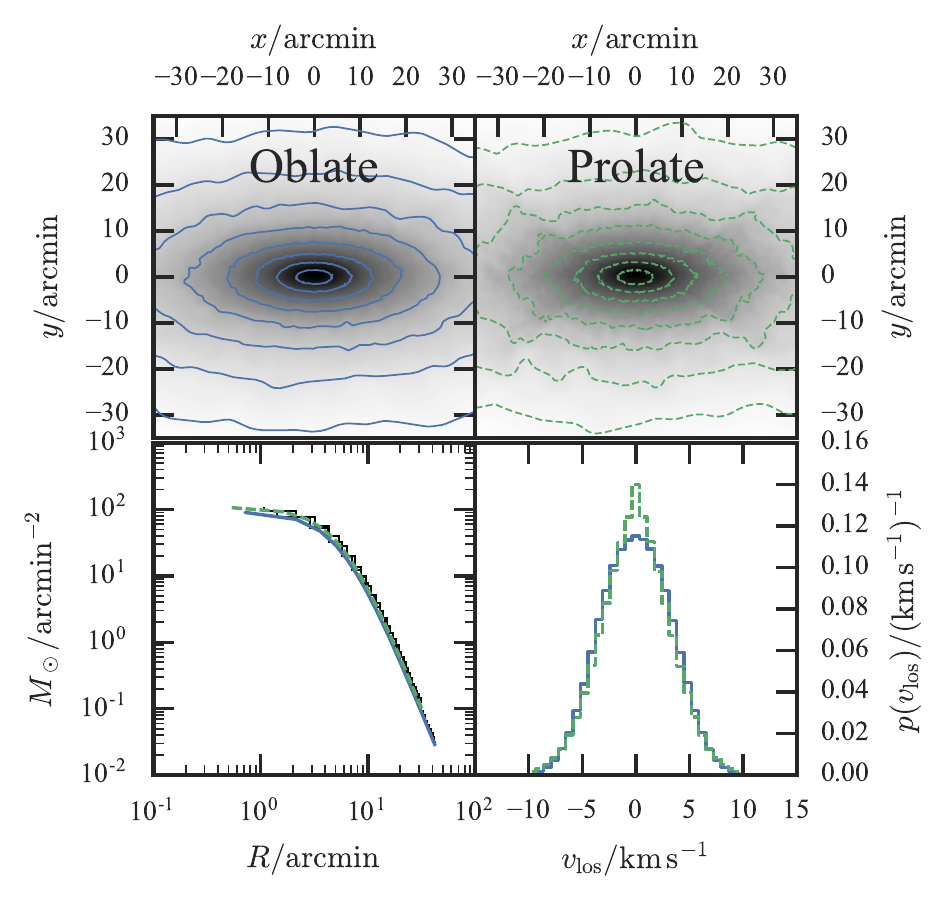}
\caption{Reticulum~II M2M equilibria of a flattened Plummer
  distribution of stars in a flattened NFW dark halo. The top left
  panel shows the logarithm of the projected mass distribution of an
  oblate model viewed edge-on with axis ratio $0.4$. The contours are
  logarithmically spaced. The top right panel shows the logarithm of
  the projected mass distribution of a prolate model viewed edge-on
  with axis ratio $0.4$. Note the `X'-shape in the prolate case. The
  bottom left panel shows the surface density profiles in elliptical
  bins with Plummer profile fits (oblate in blue, prolate in dashed
  green). The bottom right panel shows the line-of-sight velocity
  distributions (oblate in blue, prolate in dashed green).}
\label{fig:flattened}
\end{figure}

\section{Made-to-measure flattened equilibria}\label{Sect::M2M}

We begin our analysis of the J-factors of flattened dSphs with numerical models
constructed by the made-to-measure (M2M) methods \citep{Sy96} as implemented
by \citet{De09}. The models are two-component: dark and stellar. Each
component has a target density of the form
\begin{equation}
\rho (m)\propto p^{-1}q^{-1}
\Big(\frac{m}{\rs}\Big)^{-\gamma}\Big(1+\Big(\frac{m}{\rs}\Big)^{\alpha}\Big)^{(\gamma-\beta)/\alpha}\mathrm{sech}\frac{m}{\rt},
\label{abgmodel}
\end{equation}
where $m^2=x^2+(y/p)^2+(z/q)^2$. This is the familiar
double power-law with scale radius $\rs$, with an exponential taper at
the tidal radius $\rt$. For $r \ll \rs$, the density falls like
$r^{-\gamma}$, whilst for $r \gg \rs$, it falls like $r^{-\beta}$.
The case $\alpha =1, \beta =3, \gamma=1$ is the NFW dark halo.
Plummer models are often used to describe the light profiles of
dSphs~\citep[see e.g.,][]{Ir95, Ag12b}. They correspond to the
parameters $\alpha =2, \beta =5, \gamma=1$ and $\rt = \infty$.

We begin by constructing two flattened spheroidal ($p=1$) models of the Reticulum~II
dSph. For both models, the dark halo is a NFW model
($\alpha=1,\beta=3,\gamma=1,r_s=1,\rt=10$). The stars follow a Plummer
profile ($\alpha=2,\beta=5,\gamma=0,r_s=0.5,r_t=9$). The chosen ratio of the dark matter scale radius to the stellar scale radius lies within the measured range for the Local Group dSphs ($\sim1.25$ to $\sim30$) \citep{Amorisco2012}. The two models
differ in their shape. The first model is oblate in both the stars and
the dark matter with an axis ratio of $q=0.4$ (chosen to match the observed axis ratio of Reticulum~II of $0.39$ \cite{Ko15}). The second model is
prolate with an axis ratio of $q=2.5$. When viewed along the
$x$-axis both models appear flattened with axis ratio $0.4$.  In
addition, we construct a third spherical model as a reference. This
has the same parameters, but without the flattening in either the dark
matter or the stars.

The dark NFW halos source the potential (computed using a bi-orthonormal basis expansion \cite{De09}) in which the weights of the
Plummer models are adjusted until the target densities are reached. No
other constraints on the distribution functions are used. We use a
$10^7$ particle realization of the flattened NFW distribution to
compute the potential. The constraints on the Plummer model are
generated with $100$ realizations of $10^6$ particles and $10^6$
particles are used in the M2M simulation. To check convergence, the
models were run turning off the weight adjustment in the M2M
code. Both flattened models exhibit a slow drift in the density
constraint suggesting they are not perfect equilibrium
models. However, this is almost certainly true for the actual dSphs
which reside in the tidal field of the Milky Way.

Reticulum~II has a half-light major axis length of $5.63\,\mathrm{arcmin}$, is at
a distance of $\sim 30\kpc$ \citep{Ko15} and has a line-of-sight
velocity dispersion of $3.22\kms$ \citep{Ko15a}. To match the final
models to the observed constraints on Reticulum~II, we compute the
projected half-light major axis length (fitted with a Plummer model)
and the projected line-of-sight velocity dispersion. We then compute
the scale factors $\mathcal{R}$ and $\mathcal{V}$ that scale the
radial distributions and the velocity distributions to the
observations. The corresponding total mass of the dark matter profile
(set to unity in the simulation) is then scaled by a factor
$\mathcal{M}=\mathcal{R}\mathcal{V}^2$. For the spherical model we match the half-light major axis length to an `ellipticity-corrected' radius given by the geometric mean of the half-light major and minor axis lengths. This is related to the observed half-light major axis length $R_\mathrm{h}$ as $R_\mathrm{h}\sqrt{1-\epsilon}$ where $\epsilon$ is the ellipticity.

In Figs. \ref{fig:flattened}, we show the final projected
distributions of the two flattened models. Note that for the prolate
case, the models do not completely reproduce the target density profile
as there is a clear `X' shape in the $(x,y)$ plane. Additionally, we
show the surface density of the two models (using a mass-to-light
ratio of $500$, \cite{Ko15a}) and the line-of-sight velocity
distributions. The prolate velocity distribution is slightly peakier
than the oblate case but such a small difference would not be
detectable observationally.

To explore the effects of adjusting the stellar and dark-matter
profiles, we also build two further models, one with a central cusp in
the stellar profile ($\gamma=1$) and one with a cored dark-matter
profile with parameters $\alpha=1$, $\beta=4$ and $\gamma=0$.

\subsection{J- and D-factors}

For our five models of Reticulum~II, we proceed to calculate the J-
and D-factors. The J-factor for a distant source is given by\footnote{When computing these integrals numerically, we have found it useful to perform the coordinate transformation $\tan\chi = z/r_s$ where $r_s$ is the scale radius of the density profile.}
\begin{equation}
\Jf(\theta) = \frac{1}{D^2}\int_{-\infty}^{+\infty}\mathrm{d}z\,\int_0^{D\theta}\mathrm{d}R\,R\int_0^{2\pi}\mathrm{d}\phi\,\rho_\mathrm{DM}^2,
\end{equation}
where $D$ is the distance to the source ($30\kpc$ for Reticulum~II) and $\theta$ is the beam angle. Similarly, the D-factor is given by
\begin{equation}
\Df(\theta) = \frac{1}{D^2}\int_{-\infty}^{+\infty}\mathrm{d}z\,\int_0^{D\theta}\mathrm{d}R\,R\int_0^{2\pi}\mathrm{d}\phi\,\rho_\mathrm{DM}.
\end{equation}

In Table~\ref{TableRet2} we report the J- and D-factors at
$\theta=0.5^\circ$ (the typical observational resolution). We also
show the J- and D-factors for the spherical model computed from the
formulae of Paper I. We see that these formula underestimate the
J-factor by a factor $1.2$ and the D-factor by a factor $1.05$. We also
record the correction factor between the prolate / oblate models and the
spherical models using the notation
\begin{equation}
\begin{split}
\mathcal{F}_\Jf &= \log_{10}(\Jf/\Jf_\mathrm{sph}),\\
\mathcal{F}_\Df &= \log_{10}(\Df/\Df_\mathrm{sph}).
\end{split}
\label{eq:defcorrfactors}
\end{equation}
The oblate model with NFW dark matter and Plummer light has a J-factor
that is diminished by a factor $1.4$ over the spherical model and a
D-factor that is diminished by a factor $1.3$. On the other hand, the
prolate model has an enhancement in the J-factor by a factor $3.4$ and
a small decrease in the D-factor of $10\percent$. The near-prolate
model with a cuspy stellar profile produces a very similar J-factor to
the Plummer prolate model, but here the D-factor is enhanced over the
spherical model by $20\percent$. Finally, in a similar fashion to the
prolate NFW profile, the prolate cored dark matter profile also
produces an enhancement in the J-factor of a factor $3$ and a small
diminution in the D-factor of order $10\percent$.

\begin{table*}
\caption{J- and D-factors for a beam angle of $0.5^\circ$ for a series
  of Reticulum~II models. The J-factors are in units of $\mathrm{GeV^2\,cm}^{-5}$ and the D-factors are in units of $\mathrm{GeV\,cm}^{-2}$. Each model was normalized such that the
  line-of-sight velocity dispersion and half-light major axis length matched that
  of Reticulum~II. For the spherical model, an `ellipticity-corrected' half-light radius of $R_\mathrm{h}\sqrt{1-\epsilon}$ (where $\epsilon$ is the ellipticity) was used to scale the models. Note the correction factors are with respect to the spherical NFW, spherical Plummer model in the first row \emph{not} with respect to the corresponding spherical model.}
\begin{tabular}{lcccccc}
\hline
\hline
Model & Paper I J & Paper I D & $\log_{10}(\mathrm{J}(0.5^\circ))$& $\log_{10}(\mathrm{D}(0.5^\circ))$& $\mathcal{F}_\mathrm{J}$& $\mathcal{F}_\mathrm{D}$\\ 
\hline
\makecell[l]{Spherical NFW\\ Spherical Plummer}&$18.56$&$17.56$&$18.64$&$17.58$&$0.00$&$0.00$\\
\\
\makecell[l]{Oblate NFW, $p=1,\,q=0.4$\\ Oblate Plummer, $p=1,\,q=0.4$}&-&-&$18.45$&$17.62$&$-0.19$&$0.05$\\
\\
\makecell[l]{Prolate NFW, $p=1,\,q=2.5$\\ Prolate Plummer, $p=1,\,q=2.5$}&-&-&$19.05$&$17.67$&$0.40$&$0.09$\\
\\
\makecell[l]{Near-prolate NFW, $p=0.5$, $q=0.4$\\ Near-prolate cuspy Plummer\\\quad$\alpha_\star=2,\,\beta_\star=5,\,\gamma_\star=1,\,p=0.4,\,q=0.38$}&-&-&$19.01$&$17.78$&$0.37$&$0.20$\\
\\
\makecell[l]{Prolate cored DM\\\quad$\alpha_\mathrm{DM}=1,\,\beta_\mathrm{DM}=4,\,\gamma_\mathrm{DM}=0,\,p=1,\,q=2.5$\\ Prolate Plummer, $p=1,\,q=2.5$}&-&-&$18.90$&$17.66$&$0.26$&$0.08$\\
\\
\hline
\end{tabular}

\label{TableRet2}
\end{table*}

\subsection{A Range of Flattenings}

We have established that a prolate model of Reticulum~II viewed
edge-on produces a significant enhancement in the J-factor over its
spherical counterpart, whilst an oblate model has a slight
diminution. However, the observed dSphs span a whole range of
ellipticities, so we now go on to explore models with a variety of
flattenings. We construct 3 oblate M2M models with the same parameters
as the spherical reference model in Table~\ref{TableRet2} but with
flattenings $q=0.5,0.6,0.7$, and similarly 3 prolate M2M models with
flattenings $q=1.423,1.667,2$. Again the M2M models are normalized to
match the line-of-sight velocity dispersion and half-light major-axis length of
Reticulum~II.

\begin{figure}
\includegraphics[width=\columnwidth]{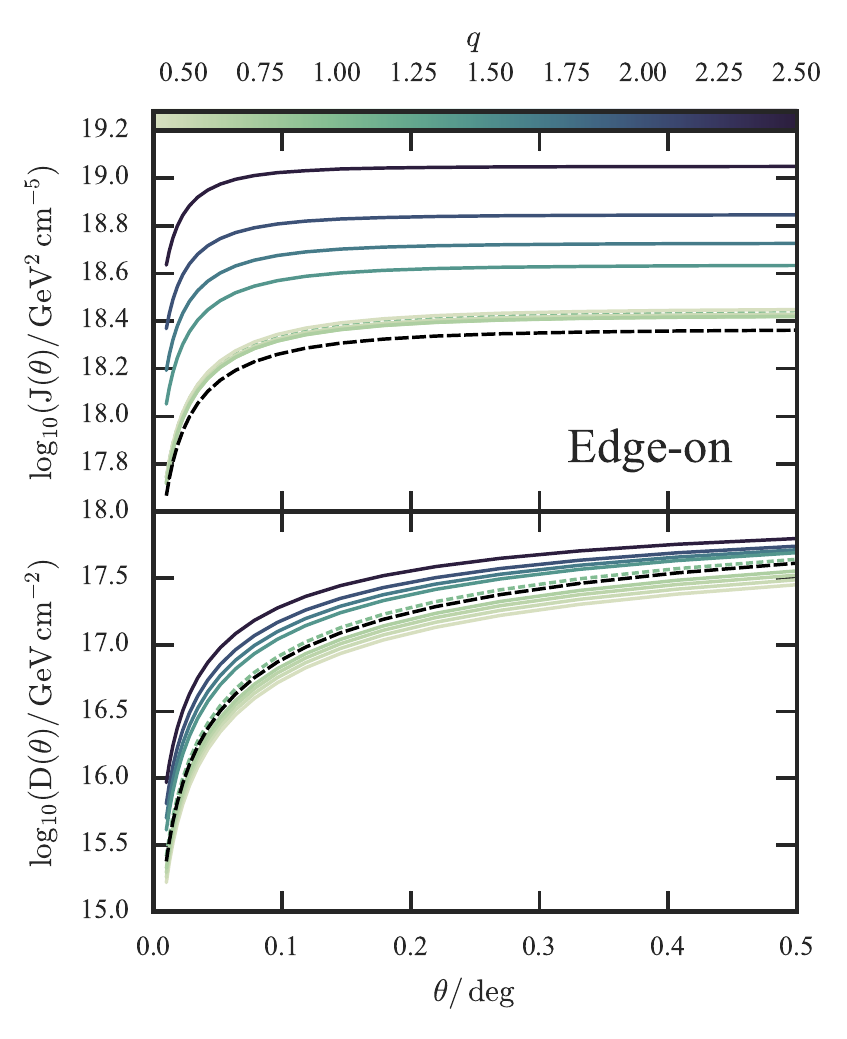}
\caption{J- and D-factors as a function of beam angle for a range of
  flattened models viewed edge-on with identical line-of-sight
  velocity dispersions and half-light major-axis lengths. The models are
  coloured by the flattening in the density of both the stars and dark matter,
  $q$. The spherical model is shown with the short-dashed line, whilst
  the analytic formula for the NFW model (equations (15) and (19) from
  Paper I) is shown with the long-dashed line.}
\label{fig:sim_profiles_edge}
\end{figure}

The J- and D-factors for our series of models are plotted in
Fig.~\ref{fig:sim_profiles_edge}. All models are viewed such that they
appear maximally flattened (along the short-axis for the prolate cases
and along the long-axis for the oblate cases). We also show the J- and
D-factors computed using the simple formulae (equations 15 and 19)
from Paper I. We see that this formula disagrees with the spherical
case by $\sim0.2$ due to the use of the empirical relation for the
half-light mass. As shown in Paper I, for most dSphs this is less than the uncertainty in the J-factor due to uncertainties in the line-of-sight velocity dispersion and half-light radius.

The prolate models produce a sequence of more enhanced J-factor at all
angles as we increase the flattening $q$. The oblate models produce a
similar sequence of decreasing J as we decrease the flattening
$q$. These trends are reproduced in the D-factor. Note the asymmetry
with $q$ in both J and D: the equivalent flattening for an prolate
model produces a larger difference from the spherical model than the corresponding oblate model.

\begin{figure}
\includegraphics[width=\columnwidth]{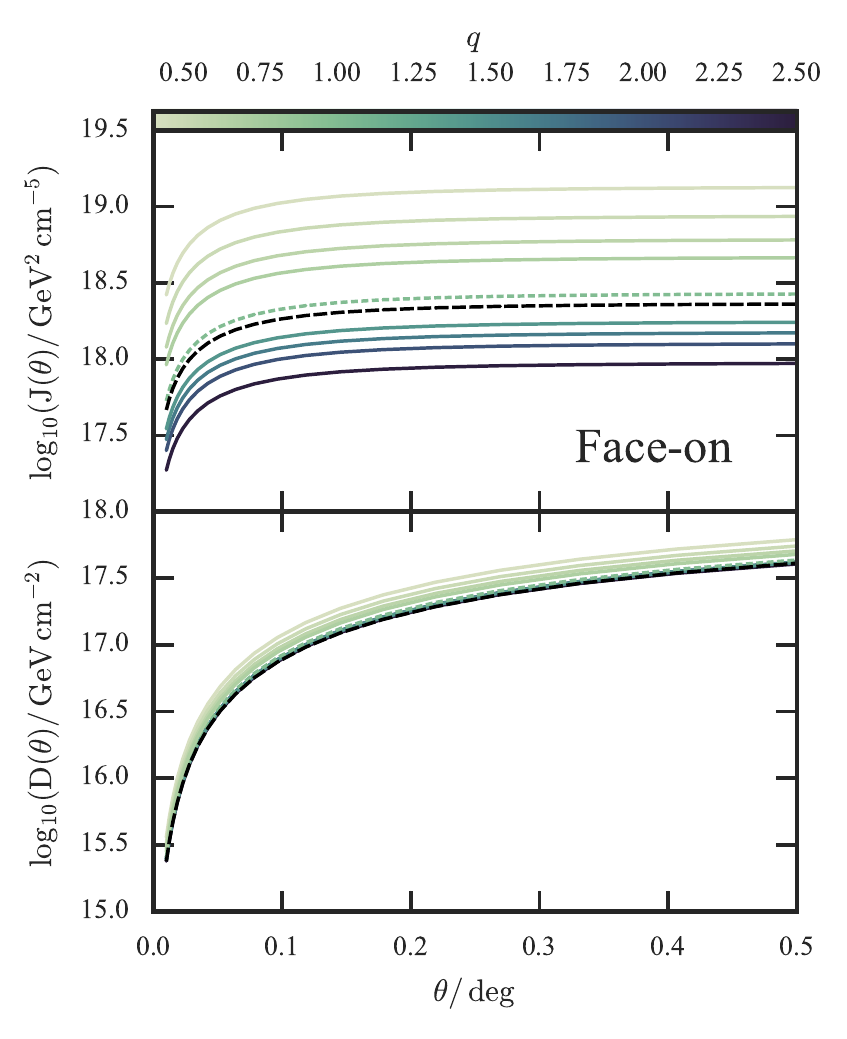}
\caption{J- and D-factors as a function of beam angle for a range of
  flattened models viewed face-on with identical line-of-sight
  velocity dispersions and half-light major-axis lengths. The models are
  coloured by the flattening in the density of both the stars and dark matter,
  $q$. The spherical model is shown with the short-dashed line and the
  model using the formulae from Paper I is shown with the long-dashed
  line.}
\label{fig:sim_profiles_round}
\end{figure}

With this sequence of models, we also investigate how the J-factor for
an apparently round dSph changes as the dSph is flattened along the
line of sight. In Fig.~\ref{fig:sim_profiles_round}, we show the range
of J- and D-factors for the set of flattened models viewed face-on
such that the isophotes appear round and all models have the same half-light radius. We see that the range of
possible J-factors with flattening along the line of sight varies by a
factor of $10$. The oblate models all have a similar decrease in the
J-factor. The D-factor is unaffected by flattening along the line of
sight.

For the series of flattened M2M models, we compute the correction factors $\mathcal{F}_\mathrm{J}$ and $\mathcal{F}_\mathrm{D}$ by comparing each model with the spherical model with the same line-of-sight velocity dispersion and the `ellipticity-corrected' half-light radius. The trends of $\mathcal{F}_{\Jf}$ and $\mathcal{F}_{\Df}$ with respect to $q$ are very smooth so
we opt to fit the corrections from the models with a simple functional
form
\begin{equation}
\mathcal{F}_{\mathrm{fit}} = \eta\log_{10}(q)
\label{eq::simple_fit}
\end{equation}
where we fit $q<1$ and $q>1$ separately. The values of $\eta$ chosen are given in
Table~\ref{table:fits}. Although our fit is an extrapolation for
$q<0.4$ and $q>2.5$, we will see that it agrees well with the more
involved models of Section~\ref{sec:AnalyticModels}. In reality, the correction factors are a function of the beam angle. We have found that the correction factors are very insensitive to the beam angle so this formula is appropriate for all dSphs irrespective of their size compared to the resolution of the instrument.

\begin{table}
\caption{Slopes $\eta$ of the base-10 logarithms of the correction factors with respect to
  $\log_{10}q$ fitted to the made-to-measure models of
  Section~\ref{Sect::M2M}. The prolate and oblate cases are treated
  separately. The multiplicative factor by which a J- or D-factor from
  a spherical analysis must be corrected is given by $q^\eta$. Note the spherical models to which we compare use an `ellipticity-corrected' half-light radius of $R_\mathrm{h}\sqrt{1-\epsilon}$ where $\epsilon=1-q$ is the ellipticity in the oblate case and $\epsilon=1-1/q$ in the prolate case.}
\begin{centering}
\begin{tabular}{lccc}
\hline
\hline
& View & $\eta$ Oblate ($q<1$) & $\eta$ Prolate ($q>1$)\\
\hline
$\mathcal{F}_\mathrm{J}$ & Edge-on&0.534&0.899\\
&Face-on&-1.647&-1.181\\
$\mathcal{F}_\mathrm{D}$ & Edge-on&0.056&0.177\\
&Face-on&-0.335&-0.089\\
\hline
\end{tabular}

\end{centering}
\label{table:fits}
\end{table}

\section{J- and D-Factors for the Milky Way dSphs}

We now apply the corrections to the J- and D-factors of the observed
dSphs. They are listed in Table~\ref{table:Jfs} along with their
measured ellipticities $\epsilon=1-b/a$ where $b/a$ is the observed
axis ratio. We take the majority of the ellipticities and $\pm1\sigma$
error-bars from the review of \cite{Mc12}. The ellipticities of the
new dSphs discovered in the Dark Energy Survey are taken from
\cite{Ko15}, the ellipticity of Pisces II is taken from \cite{Be10}
and that of Hydra II from \cite{Ma15}. For both Leo T and Horologium
I, only upper-bounds on the ellipticity are available.

For each dSph, we compute the correction factor assuming the dSph is
either oblate or prolate and observed edge-on. We draw samples from
the error distributions of the ellipticities and compute the median
and $\pm1\sigma$ values of the correction factors for both the J- and
D-factors using equation~\eqref{eq::simple_fit}. The baseline spherical model to which we are comparing uses an effective half-light radius of $R_\mathrm{h}\sqrt{1-\epsilon}$. We combine these estimates with the spherical estimates
computed in Paper I (adding the errors in quadrature). The results of
this procedure are reported in Tables~\ref{table:Jfs} and
\ref{table:Dfs}.

\begin{figure*}
$$\includegraphics[width=\textwidth]{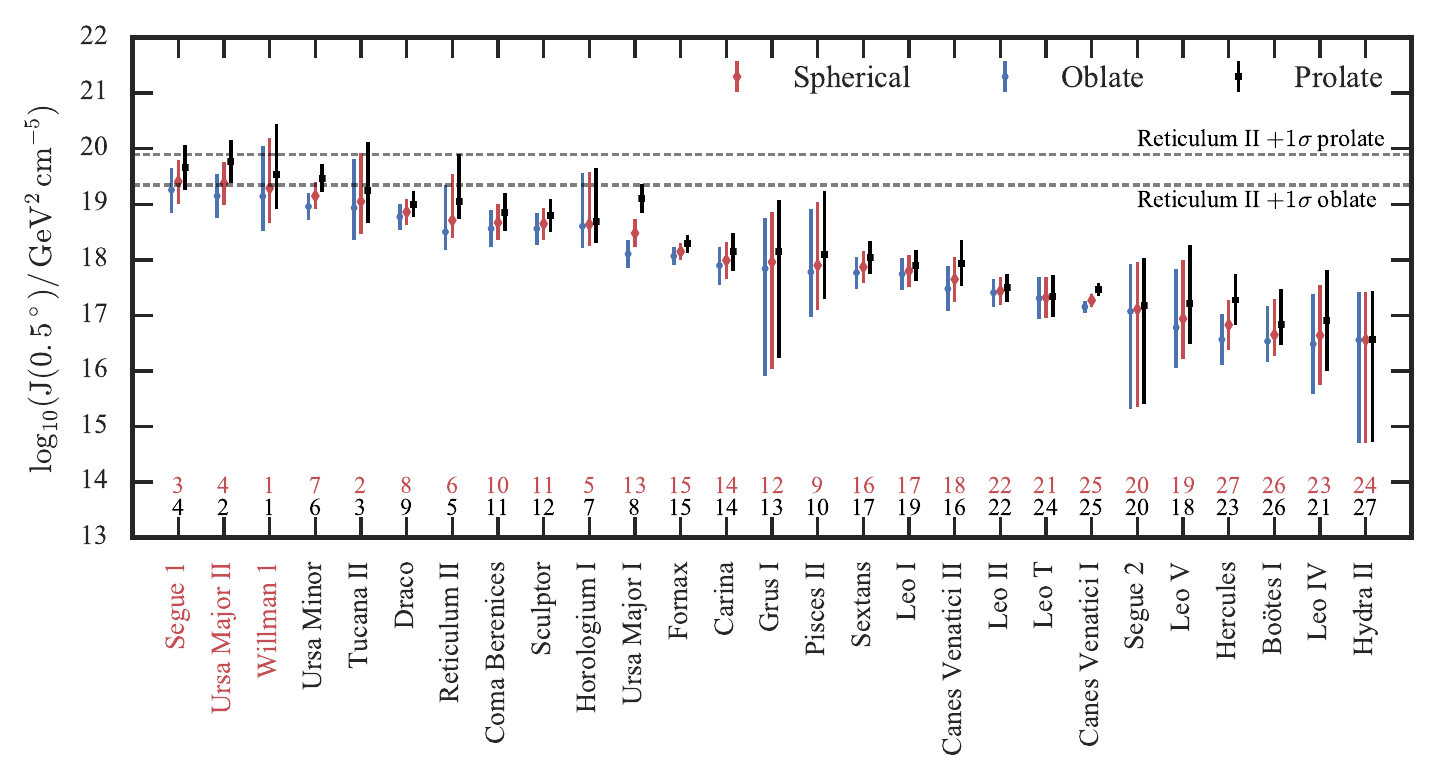}$$
$$\includegraphics[width=\textwidth]{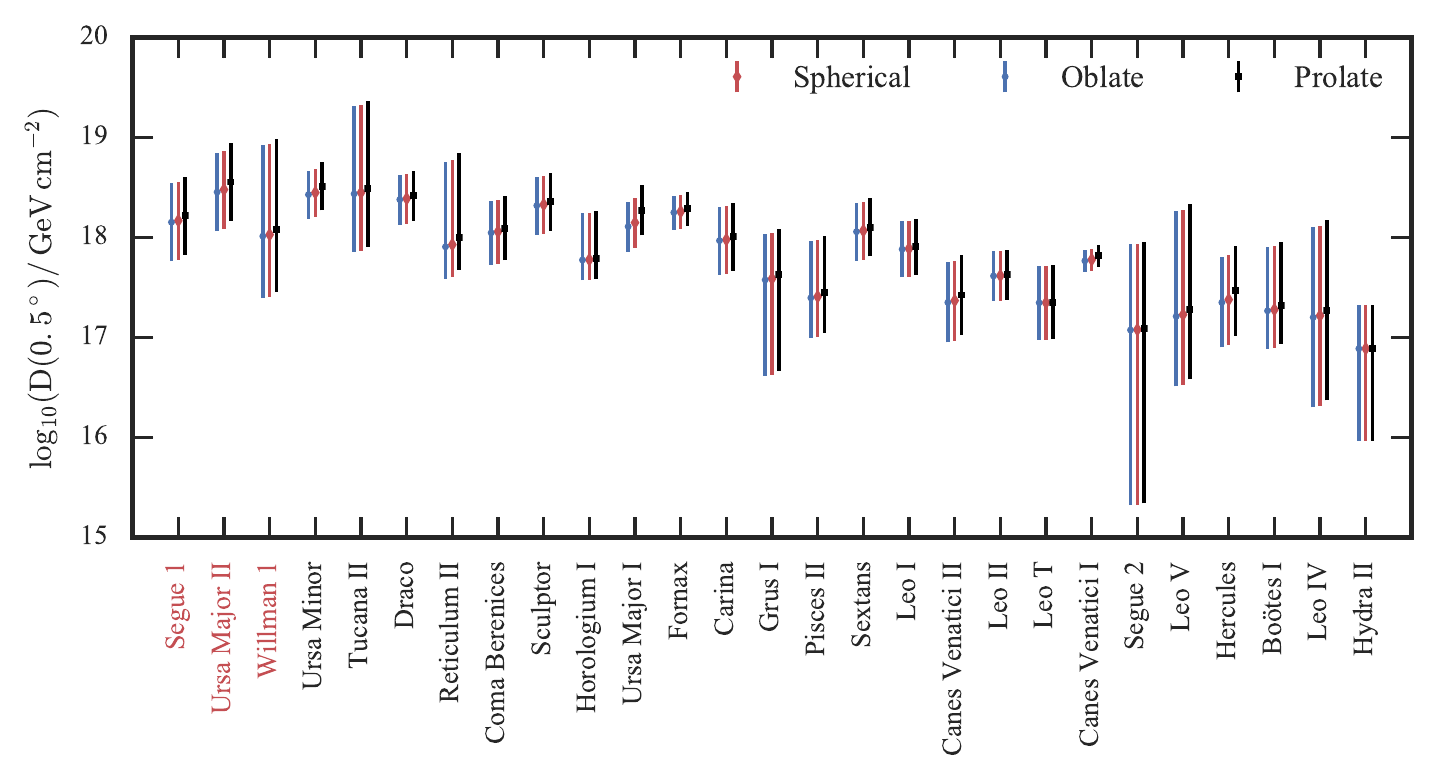}$$
\caption{J- (top) and D-factors (bottom) integrated over a beam angle
  of $0.5^\circ$ for $27$ dSphs. The diamonds with red error-bars are
  computed assuming a spherical model and are taken from Paper
  I. The circles with blue error-bars show the spherical J-factors adjusted by the
  oblate correction factors marginalized over the uncertainty in the
  ellipticity (assuming the galaxy is observed edge-on) and the squares with black
  error-bars show the spherical J-factors adjusted by the prolate
  correction factors marginalized over the uncertainty in the
  ellipticity (assuming the galaxy is observed edge-on). The dSphs
  are ordered by their median spherical J-factors. The top set of red
  numbers gives the ordering of the upper-limits on the spherical
  J-factors, and the bottom set of black numbers gives the ordering of
  the upper-limits on the prolate J-factors. The gray dashed lines
  show the $1\sigma$ upper-limit for the Reticulum~II assuming it is
  prolate or oblate. The three dSphs with red names have unknown additional systematic uncertainties due to the presence of contaminants or the questionable assumption of dynamical equilibrium.}
\label{fig:KF_corrections}
\end{figure*}

We show this data in Fig.~\ref{fig:KF_corrections}. We plot the
distribution of J- and D-factors for the dSphs assuming they are
spherical, oblate or prolate. Ursa Major I has the largest ellipticity
and hence the largest prolate correction factor (a factor of
$\sim4$). Reticulum~II, Ursa Major II and Hercules all have ellipticities
$\sim0.6$ and so the prolate correction factors are approximately
$\sim2.2-2.7$. For ellipticities less than $\sim0.4$, the correction
factors are less than the errors on the spherical J-factors. For every
dSph the correction to the D-factors is smaller than the errors in
the spherical D-factor. Hence, we conclude that flattening has a
negligible effect on the D-factor estimates.

If the entire population of dSphs is prolate then only Tucana II and
Willman 1 have potentially higher J-factors than Reticulum~II, with
both Ursa Major II and Segue 1 having a very similar J-factor to
Reticulum~II. We remark that Tucana II is consistent with having
circular isophotes \cite{Ko15}, whilst the assumption of dynamical
equilibrium for Willman 1 is dubious \cite{Willman2011}. Similarly, Ursa Major II appears
to be in process of severe tidal disruption \citep{Zu06}. Finally, the
J-factor of Segue 1 has been shown to be extremely sensitive to the presence of foreground
contaminants \citep[e.g.,][]{NO09,Bo15c}. These final three dSphs have been marked in red in Figure~\ref{fig:KF_corrections} to indicate their dubious J-factors. Therefore, it is possible that the Reticulum~II gamma-ray signal may be due to annihilation if the dwarf has a prolate shape. We can robustly conclude from Figure~\ref{fig:KF_corrections} that if Reticulum~II has a prolate shape then an observed annihilation signal from only Reticulum~II is not in tension with the lack of signals from all the other dSphs irrespective of their shapes. If, however, Reticulum~II is oblate and has an observed annihilation signal we begin to have some tension if there is a lack of signal from the other dSphs. The majority of this tension arises from those problematic dSphs already mentioned. However, if both Ursa Minor and Tucana II have prolate shapes it becomes unlikely that they both have smaller J factors than an oblate Reticulum II.

\begin{turnpage}
\begin{table*}
\caption{Annihilation correction factors for dwarf spheroidals due to
  their observed ellipticity $\epsilon$ (note Leo T and Horologium I
  only have upper-bounds on the ellipticity). We report the spherical
  J-factor for a beam angle of $0.5^\circ$ in units of $\mathrm{GeV^2\,cm}^{-5}$ along with the corrections $\mathcal{F}_\mathrm{J}$
  assuming the galaxy is observed exactly edge-on and is either oblate
  or prolate. We report the resultant J-factors for these cases as
  $\Jf_\mathrm{obl}$ and $\Jf_\mathrm{pro}$ in units of
  $\mathrm{GeV^2\,cm^{-5}}$. The dSphs are ordered by their
  ellipticity.}
\begin{centering}
\begin{tabular}{lc|ccccc}
\hline
\hline
Name & $\epsilon$ & $\log_{10}(\mathrm{J}_\mathrm{sph}(0.5^\circ))$ & Oblate $\mathcal{F}_\mathrm{J}$ & Prolate $\mathcal{F}_\mathrm{J}$ & $\log_{10}(\mathrm{J}_\mathrm{obl}(0.5^\circ))$ & $\log_{10}(\mathrm{J}_\mathrm{pro}(0.5^\circ))$\\ 
\hline
Hydra II&$0.01_{-0.01}^{+0.20}$&$16.56_{-1.85}^{+0.87}$&$-0.002_{-0.052}^{+0.002}$&$0.004_{-0.003}^{+0.087}$&$16.56_{-1.85}^{+0.87}$&$16.56_{-1.85}^{+0.87}$\\
Leo T&$<0.10$&$17.32_{-0.37}^{+0.38}$&$-0.012_{-0.009}^{+0.008}$&$0.020_{-0.014}^{+0.015}$&$17.31_{-0.37}^{+0.38}$&$17.34_{-0.37}^{+0.38}$\\
Leo II&$0.13_{-0.05}^{+0.05}$&$17.44_{-0.25}^{+0.25}$&$-0.032_{-0.017}^{+0.011}$&$0.054_{-0.018}^{+0.029}$&$17.41_{-0.25}^{+0.25}$&$17.49_{-0.25}^{+0.25}$\\
Segue 2&$0.15_{-0.10}^{+0.10}$&$17.11_{-1.76}^{+0.85}$&$-0.037_{-0.042}^{+0.019}$&$0.063_{-0.032}^{+0.070}$&$17.07_{-1.76}^{+0.85}$&$17.17_{-1.76}^{+0.85}$\\
Leo I&$0.21_{-0.03}^{+0.03}$&$17.80_{-0.28}^{+0.28}$&$-0.055_{-0.010}^{+0.008}$&$0.092_{-0.014}^{+0.017}$&$17.75_{-0.28}^{+0.28}$&$17.89_{-0.28}^{+0.28}$\\
Horologium I&$<0.28$&$18.64_{-0.39}^{+0.95}$&$-0.034_{-0.028}^{+0.024}$&$0.058_{-0.040}^{+0.047}$&$18.61_{-0.39}^{+0.95}$&$18.70_{-0.39}^{+0.95}$\\
Fornax&$0.30_{-0.01}^{+0.01}$&$18.15_{-0.16}^{+0.16}$&$-0.083_{-0.003}^{+0.003}$&$0.139_{-0.005}^{+0.006}$&$18.07_{-0.16}^{+0.16}$&$18.29_{-0.16}^{+0.16}$\\
Draco&$0.31_{-0.02}^{+0.02}$&$18.86_{-0.24}^{+0.24}$&$-0.086_{-0.007}^{+0.006}$&$0.145_{-0.011}^{+0.012}$&$18.77_{-0.24}^{+0.24}$&$19.00_{-0.24}^{+0.24}$\\
Sculptor&$0.32_{-0.03}^{+0.03}$&$18.65_{-0.29}^{+0.29}$&$-0.090_{-0.011}^{+0.010}$&$0.151_{-0.016}^{+0.019}$&$18.56_{-0.29}^{+0.29}$&$18.80_{-0.29}^{+0.29}$\\
Carina&$0.33_{-0.05}^{+0.05}$&$17.99_{-0.34}^{+0.34}$&$-0.093_{-0.020}^{+0.016}$&$0.156_{-0.026}^{+0.034}$&$17.90_{-0.34}^{+0.34}$&$18.15_{-0.34}^{+0.34}$\\
Sextans&$0.35_{-0.05}^{+0.05}$&$17.87_{-0.29}^{+0.29}$&$-0.100_{-0.020}^{+0.016}$&$0.169_{-0.028}^{+0.034}$&$17.77_{-0.29}^{+0.29}$&$18.04_{-0.29}^{+0.29}$\\
Coma Berenices&$0.38_{-0.14}^{+0.14}$&$18.67_{-0.32}^{+0.33}$&$-0.108_{-0.071}^{+0.039}$&$0.182_{-0.066}^{+0.120}$&$18.56_{-0.33}^{+0.33}$&$18.85_{-0.33}^{+0.35}$\\
Bo\"otes I&$0.39_{-0.06}^{+0.06}$&$16.65_{-0.38}^{+0.64}$&$-0.114_{-0.026}^{+0.020}$&$0.192_{-0.033}^{+0.043}$&$16.54_{-0.38}^{+0.64}$&$16.84_{-0.38}^{+0.64}$\\
Canes Venatici I&$0.39_{-0.03}^{+0.03}$&$17.27_{-0.11}^{+0.11}$&$-0.115_{-0.012}^{+0.011}$&$0.194_{-0.018}^{+0.020}$&$17.15_{-0.11}^{+0.11}$&$17.46_{-0.11}^{+0.11}$\\
Tucana II&$0.39_{-0.20}^{+0.10}$&$19.05_{-0.58}^{+0.87}$&$-0.114_{-0.049}^{+0.053}$&$0.193_{-0.090}^{+0.082}$&$18.94_{-0.58}^{+0.87}$&$19.24_{-0.59}^{+0.87}$\\
Pisces II&$0.40_{-0.10}^{+0.10}$&$17.90_{-0.80}^{+1.14}$&$-0.119_{-0.048}^{+0.032}$&$0.200_{-0.053}^{+0.081}$&$17.78_{-0.80}^{+1.14}$&$18.10_{-0.80}^{+1.14}$\\
Grus I&$0.41_{-0.28}^{+0.20}$&$17.96_{-1.93}^{+0.90}$&$-0.117_{-0.117}^{+0.064}$&$0.198_{-0.107}^{+0.196}$&$17.84_{-1.93}^{+0.90}$&$18.16_{-1.93}^{+0.92}$\\
Willman 1&$0.47_{-0.08}^{+0.08}$&$19.29_{-0.62}^{+0.91}$&$-0.147_{-0.040}^{+0.030}$&$0.247_{-0.051}^{+0.067}$&$19.14_{-0.62}^{+0.91}$&$19.54_{-0.62}^{+0.91}$\\
Segue 1&$0.48_{-0.13}^{+0.13}$&$19.41_{-0.40}^{+0.39}$&$-0.151_{-0.076}^{+0.045}$&$0.255_{-0.076}^{+0.128}$&$19.26_{-0.41}^{+0.39}$&$19.66_{-0.41}^{+0.41}$\\
Leo IV&$0.49_{-0.11}^{+0.11}$&$16.64_{-0.90}^{+0.90}$&$-0.156_{-0.064}^{+0.041}$&$0.262_{-0.068}^{+0.108}$&$16.48_{-0.90}^{+0.90}$&$16.90_{-0.90}^{+0.91}$\\
Leo V&$0.50_{-0.15}^{+0.15}$&$16.94_{-0.72}^{+1.05}$&$-0.159_{-0.096}^{+0.053}$&$0.268_{-0.089}^{+0.161}$&$16.78_{-0.73}^{+1.05}$&$17.21_{-0.73}^{+1.06}$\\
Canes Venatici II&$0.52_{-0.11}^{+0.11}$&$17.65_{-0.40}^{+0.40}$&$-0.171_{-0.065}^{+0.042}$&$0.288_{-0.071}^{+0.109}$&$17.48_{-0.41}^{+0.40}$&$17.94_{-0.41}^{+0.41}$\\
Ursa Minor&$0.56_{-0.05}^{+0.05}$&$19.15_{-0.24}^{+0.25}$&$-0.191_{-0.029}^{+0.024}$&$0.321_{-0.040}^{+0.049}$&$18.96_{-0.24}^{+0.25}$&$19.47_{-0.24}^{+0.25}$\\
Reticulum II&$0.59_{-0.03}^{+0.02}$&$18.71_{-0.32}^{+0.84}$&$-0.207_{-0.012}^{+0.016}$&$0.348_{-0.027}^{+0.020}$&$18.50_{-0.32}^{+0.84}$&$19.06_{-0.32}^{+0.84}$\\
Ursa Major II&$0.63_{-0.05}^{+0.05}$&$19.38_{-0.39}^{+0.39}$&$-0.231_{-0.035}^{+0.028}$&$0.389_{-0.047}^{+0.060}$&$19.15_{-0.39}^{+0.39}$&$19.77_{-0.39}^{+0.39}$\\
Hercules&$0.68_{-0.08}^{+0.08}$&$16.83_{-0.45}^{+0.45}$&$-0.266_{-0.069}^{+0.050}$&$0.447_{-0.084}^{+0.116}$&$16.56_{-0.46}^{+0.45}$&$17.28_{-0.46}^{+0.46}$\\
Ursa Major I&$0.80_{-0.04}^{+0.04}$&$18.48_{-0.25}^{+0.25}$&$-0.373_{-0.053}^{+0.041}$&$0.629_{-0.070}^{+0.089}$&$18.11_{-0.26}^{+0.25}$&$19.11_{-0.26}^{+0.27}$\\
\hline
\end{tabular}

\end{centering}
\label{table:Jfs}
\end{table*}
\end{turnpage}

\begin{turnpage}
\begin{table*}
\caption{As Table~\ref{table:Jfs}, but for the decay correction
  factors. The D-factors are quoted in units of $\mathrm{GeV\,cm}^{-2}$.}
\begin{centering}
\begin{tabular}{lc|ccccc}
\hline
\hline
Name & $\epsilon$ & $\log_{10}(\mathrm{D}_\mathrm{sph}(0.5^\circ))$ & Oblate $\mathcal{F}_\mathrm{D}$ & Prolate $\mathcal{F}_\mathrm{D}$ & $\log_{10}(\mathrm{D}_\mathrm{obl}(0.5^\circ))$ & $\log_{10}(\mathrm{D}_\mathrm{pro}(0.5^\circ))$\\ 
\hline
Hydra II&$0.01_{-0.01}^{+0.20}$&$16.89_{-0.92}^{+0.44}$&$-0.000_{-0.005}^{+0.000}$&$0.001_{-0.001}^{+0.017}$&$16.89_{-0.92}^{+0.44}$&$16.89_{-0.92}^{+0.44}$\\
Leo T&$<0.10$&$17.35_{-0.37}^{+0.37}$&$-0.001_{-0.001}^{+0.001}$&$0.004_{-0.003}^{+0.003}$&$17.35_{-0.37}^{+0.37}$&$17.35_{-0.37}^{+0.37}$\\
Leo II&$0.13_{-0.05}^{+0.05}$&$17.62_{-0.25}^{+0.25}$&$-0.003_{-0.002}^{+0.001}$&$0.011_{-0.004}^{+0.006}$&$17.62_{-0.25}^{+0.25}$&$17.63_{-0.25}^{+0.25}$\\
Segue 2&$0.15_{-0.10}^{+0.10}$&$17.08_{-1.75}^{+0.86}$&$-0.004_{-0.004}^{+0.002}$&$0.012_{-0.006}^{+0.014}$&$17.08_{-1.75}^{+0.86}$&$17.09_{-1.75}^{+0.86}$\\
Leo I&$0.21_{-0.03}^{+0.03}$&$17.89_{-0.28}^{+0.28}$&$-0.006_{-0.001}^{+0.001}$&$0.018_{-0.003}^{+0.003}$&$17.88_{-0.28}^{+0.28}$&$17.91_{-0.28}^{+0.28}$\\
Horologium I&$<0.28$&$17.78_{-0.20}^{+0.47}$&$-0.004_{-0.003}^{+0.002}$&$0.011_{-0.008}^{+0.009}$&$17.78_{-0.20}^{+0.47}$&$17.79_{-0.20}^{+0.47}$\\
Fornax&$0.30_{-0.01}^{+0.01}$&$18.26_{-0.17}^{+0.17}$&$-0.009_{-0.000}^{+0.000}$&$0.027_{-0.001}^{+0.001}$&$18.25_{-0.17}^{+0.17}$&$18.29_{-0.17}^{+0.17}$\\
Draco&$0.31_{-0.02}^{+0.02}$&$18.39_{-0.25}^{+0.25}$&$-0.009_{-0.001}^{+0.001}$&$0.029_{-0.002}^{+0.002}$&$18.38_{-0.25}^{+0.25}$&$18.42_{-0.25}^{+0.25}$\\
Sculptor&$0.32_{-0.03}^{+0.03}$&$18.33_{-0.29}^{+0.29}$&$-0.009_{-0.001}^{+0.001}$&$0.030_{-0.003}^{+0.004}$&$18.32_{-0.29}^{+0.29}$&$18.36_{-0.29}^{+0.29}$\\
Carina&$0.33_{-0.05}^{+0.05}$&$17.98_{-0.34}^{+0.34}$&$-0.010_{-0.002}^{+0.002}$&$0.031_{-0.005}^{+0.007}$&$17.97_{-0.34}^{+0.34}$&$18.01_{-0.34}^{+0.34}$\\
Sextans&$0.35_{-0.05}^{+0.05}$&$18.07_{-0.29}^{+0.29}$&$-0.011_{-0.002}^{+0.002}$&$0.033_{-0.005}^{+0.007}$&$18.06_{-0.29}^{+0.29}$&$18.10_{-0.29}^{+0.29}$\\
Coma Berenices&$0.38_{-0.14}^{+0.14}$&$18.06_{-0.32}^{+0.32}$&$-0.011_{-0.007}^{+0.004}$&$0.036_{-0.013}^{+0.024}$&$18.05_{-0.32}^{+0.32}$&$18.10_{-0.32}^{+0.32}$\\
Bo\"otes I&$0.39_{-0.06}^{+0.06}$&$17.28_{-0.38}^{+0.64}$&$-0.012_{-0.003}^{+0.002}$&$0.038_{-0.007}^{+0.008}$&$17.27_{-0.38}^{+0.64}$&$17.32_{-0.38}^{+0.64}$\\
Canes Venatici I&$0.39_{-0.03}^{+0.03}$&$17.78_{-0.11}^{+0.11}$&$-0.012_{-0.001}^{+0.001}$&$0.038_{-0.004}^{+0.004}$&$17.77_{-0.11}^{+0.11}$&$17.82_{-0.11}^{+0.11}$\\
Tucana II&$0.39_{-0.20}^{+0.10}$&$18.45_{-0.58}^{+0.88}$&$-0.012_{-0.005}^{+0.006}$&$0.038_{-0.018}^{+0.016}$&$18.44_{-0.58}^{+0.88}$&$18.49_{-0.58}^{+0.88}$\\
Pisces II&$0.40_{-0.10}^{+0.10}$&$17.41_{-0.40}^{+0.57}$&$-0.012_{-0.005}^{+0.003}$&$0.039_{-0.010}^{+0.016}$&$17.40_{-0.40}^{+0.57}$&$17.45_{-0.40}^{+0.57}$\\
Grus I&$0.41_{-0.28}^{+0.20}$&$17.59_{-0.96}^{+0.46}$&$-0.012_{-0.012}^{+0.007}$&$0.039_{-0.021}^{+0.039}$&$17.58_{-0.96}^{+0.46}$&$17.63_{-0.96}^{+0.46}$\\
Willman 1&$0.47_{-0.08}^{+0.08}$&$18.03_{-0.62}^{+0.91}$&$-0.015_{-0.004}^{+0.003}$&$0.049_{-0.010}^{+0.013}$&$18.01_{-0.62}^{+0.91}$&$18.08_{-0.62}^{+0.91}$\\
Segue 1&$0.48_{-0.13}^{+0.13}$&$18.17_{-0.39}^{+0.39}$&$-0.016_{-0.008}^{+0.005}$&$0.050_{-0.015}^{+0.025}$&$18.15_{-0.39}^{+0.39}$&$18.22_{-0.39}^{+0.39}$\\
Leo IV&$0.49_{-0.11}^{+0.11}$&$17.22_{-0.90}^{+0.90}$&$-0.016_{-0.007}^{+0.004}$&$0.052_{-0.013}^{+0.021}$&$17.20_{-0.90}^{+0.90}$&$17.27_{-0.90}^{+0.90}$\\
Leo V&$0.50_{-0.15}^{+0.15}$&$17.23_{-0.70}^{+1.05}$&$-0.017_{-0.010}^{+0.006}$&$0.053_{-0.018}^{+0.032}$&$17.21_{-0.70}^{+1.05}$&$17.28_{-0.70}^{+1.05}$\\
Canes Venatici II&$0.52_{-0.11}^{+0.11}$&$17.37_{-0.40}^{+0.40}$&$-0.018_{-0.007}^{+0.004}$&$0.057_{-0.014}^{+0.022}$&$17.35_{-0.40}^{+0.40}$&$17.43_{-0.40}^{+0.40}$\\
Ursa Minor&$0.56_{-0.05}^{+0.05}$&$18.45_{-0.24}^{+0.24}$&$-0.020_{-0.003}^{+0.002}$&$0.063_{-0.008}^{+0.010}$&$18.43_{-0.24}^{+0.24}$&$18.51_{-0.24}^{+0.24}$\\
Reticulum II&$0.59_{-0.03}^{+0.02}$&$17.93_{-0.32}^{+0.85}$&$-0.022_{-0.001}^{+0.002}$&$0.069_{-0.005}^{+0.004}$&$17.91_{-0.32}^{+0.85}$&$18.00_{-0.32}^{+0.85}$\\
Ursa Major II&$0.63_{-0.05}^{+0.05}$&$18.48_{-0.39}^{+0.39}$&$-0.024_{-0.004}^{+0.003}$&$0.077_{-0.009}^{+0.012}$&$18.46_{-0.39}^{+0.39}$&$18.56_{-0.39}^{+0.39}$\\
Hercules&$0.68_{-0.08}^{+0.08}$&$17.38_{-0.45}^{+0.45}$&$-0.028_{-0.007}^{+0.005}$&$0.088_{-0.017}^{+0.023}$&$17.35_{-0.45}^{+0.45}$&$17.47_{-0.45}^{+0.45}$\\
Ursa Major I&$0.80_{-0.04}^{+0.04}$&$18.15_{-0.25}^{+0.25}$&$-0.039_{-0.006}^{+0.004}$&$0.124_{-0.014}^{+0.018}$&$18.11_{-0.25}^{+0.25}$&$18.27_{-0.25}^{+0.25}$\\
\hline
\end{tabular}

\end{centering}
\label{table:Dfs}
\end{table*}
\end{turnpage}

\section{Semi-analytic models}
\label{sec:AnalyticModels}

Numerical M2M models provide a robust method for determining the
corrections required when modeling flattened systems as
spherical. However, they are computationally expensive to construct so
cannot be employed in a Markov Chain Monte Carlo analysis that
requires many models. We have provided a simple fitting formula for
our model setup, but there will be some variation in the correction
factors depending on, for instance, the light profile,
the density profile of the dark matter, and the ratio of the scale lengths
of the light to the dark matter.

We now proceed to understand and reproduce the results of the M2M
models using simpler methods. In sub-section A, we describe a general
virial method to compute J-factors for flattened halo models. This
numerical algorithm can be applied to any dark matter density, but in
the two following subsections, we provide analytic shortcuts to
evaluate the J-factors for two specific families -- the infinite
flattened cusps and the flat rotation curve halos.  Readers primarily
interested in the results, rather than the details of the methods,
should skip to sub-section D, where we compare our models to the M2M
results. Figures~\ref{fig:corrections} and
\ref{fig:D_corrections} provide summary plots, which show the range of
correction factors as a function of the flattening of the stellar
density.

\subsection{The Virial Method}\label{VirMethod}

We can construct approximate equilibrium models much more cheaply than
with the full M2M apparatus by using the virial theorem. The two constraints
provided by the data are the integrated line-of-sight velocity
dispersion ${\langle\sigma_\mathrm{los}^2\rangle}$ and the half-light major-axis length $R_h$. We describe a method to match these
observations given density models for the light ($\rho_\star$) and dark matter ($\rho_{\mathrm{DM}}$).

\begin{enumerate}
\item For a given viewing angle $(\vartheta,\varphi)$, we find the
  measured ellipticity and orientation of the observed minor axis  (using, for instance, equations (A1,A2,A6) of \cite{Weijmans2014})  and
  compute the elliptical half-light radius $R_h'$. This gives us a
  length scaling $\mathcal{R} = R_h/R_h'$, and encodes the geometric factor described in the introduction.

\item In principle, to solve for the kinematics of the stars in the
  dark matter potential, we could use the axisymmetric Jeans
  equations. However, there are degeneracies in the solution space and
  only a few algorithms exist for solution~\citep{Ca08,Ev15}. As we need only
  match an integrated quantity, we use the virial theorem to compute $\langle\sigma_\mathrm{los}^2\rangle$ as
\begin{equation}
\langle\sigma_\mathrm{los}'^2\rangle ={W_{\rm los} \over W} \langle\sigma_\mathrm{tot}^2\rangle = {W_{\rm los} \over M},
\end{equation}
where
\begin{equation}
W_{\rm los} = \int\mathrm{d}^3\boldsymbol{x}\,\rho_\star R_{ij} x_j
\frac{\partial\Phi_{\mathrm{DM}}} {\partial x_k} R_{ki}.
\end{equation}
$\Phi_{\mathrm{DM}}$ is the dark matter potential (generically computed using a multipole expansion \cite{BT}), $M$ is the total dark matter mass and $R_{ij}$ is the projection matrix along the line-of-sight from coordinates aligned with the principal axes of the dSph. We have used the Einstein summation convention. For triaxial symmetry, the cross-terms in the integral vanish so we need only project the velocity dispersions along the principal axes. This gives us a
  velocity scaling $\mathcal{V} = \langle\sigma_\mathrm{los}^2\rangle/\langle\sigma_\mathrm{los}'^2\rangle$, and encodes the kinematic factor described in the introduction.

\item We compute a mass scaling $\mathcal{M}=\mathcal{V}^2\mathcal{R}$. The initial
  model is scaled by $\mathcal{M}$ and $\mathcal{R}$ and the J- and
  D-factors are computed. These can be compared to the spherical model
  with the same line-of-sight velocity dispersion and half-light
  radius.

\end{enumerate}

This algorithm is completely general. For some special choices of
stellar and dark matter density, the integration in the virial theorem
can be performed analytically. We now give two examples -- infinite
flattened cusps and flat rotation curve halos -- for which the virial
integrals can be done. This means that the behavior of the J-factor
at fixed observables (line of sight velocity dispersion and half-light
radii) can be mapped out analytically as a function of flattening or
concentration.

\subsection{Flattened Cusps}\label{AxiCusps}

Let us take the dark matter halo as an axisymmetric cusp
stratified on similar concentric spheroids with an axis ratio $q$. If
the cusps have the same mass $\Mh$ within the spheroidal half-light
radius $\mh$, then the mass enclosed is
\begin{equation}
M(m) = \Mh \left( {m\over \mh} \right)^{3-\gamma_\mathrm{DM}}\quad \mathrm{for } \quad m\leq r_t
\end{equation}
and $M=\Mh(r_t/\mh)^{3-\gamma_\mathrm{DM}}$ otherwise. $m^2 = x^2 + y^2 + z^2q^{-2} = R^2 + z^2q^{-2}$ and $r_t$ is a hard truncation ellipsoidal radius. The dark matter
density is
\begin{equation}
\rhoDM(m) = {\Mh\over 4 \pi q\mh^{3-\gamma_\mathrm{DM}}}{3 -\gamma_\mathrm{DM} \over m^{\gamma_\mathrm{DM}}}\quad \mathrm{for}\quad m\leq r_t,
\label{AxiCusp}
\end{equation}
and zero otherwise.
Note the factor of $q$ in the denominator which comes from the
Jacobian. It means that the oblate models ($q<1$) in the sequence have
an increased density as compared to their spherical progenitor, whilst
the prolate models ($q>1$) have a decreased density.  The spherical
member of the family obeys the empirical law (\ref{eq:wolf}).  As the
mass $\Mh$ is preserved along the sequence, we can still use
Eq.
 (\ref{eq:wolf}) for the flattened cusps provided we correct the
observables -- the line of sight velocity dispersion and the projected
half-light radius -- to the spherical parent.

For comparison purposes, it is useful to define the J-factor and
D-factor of the infinite spherical cusp ($r_t\rightarrow\infty$, equations (8) and (11) in Paper I) as
\begin{equation}
\Jf_{\rm sph} =   \frac{1}{D^2\rh^3}\Big(\frac{\langle\sigma_{\rm los}^2\rangle\rh}{G}\Big)^2\Big({ D\theta\over \rh}\Big)^{3-2\gamma_\mathrm{DM}} P(\gamma_\mathrm{DM}),
\end{equation}
\begin{equation}
\Df_{\rm sph} =   \frac{1}{D^2}\frac{\langle\sigma_{\rm los}^2\rangle\rh}{G}\Big({ D\theta\over \rh}\Big)^{3-\gamma_\mathrm{DM}} Q(\gamma_\mathrm{DM}),
\end{equation}
where both $P(\gamma_\mathrm{DM})$ and $Q(\gamma_\mathrm{DM})$ are constants given in Paper
I. In these expressions, the half-light radius $\rh$ is the `ellipticity-corrected' half-light radius that includes a factor of $\sqrt{1-\epsilon}$. The J- and D- factors for our axisymmetric models can now be
written in the form
\begin{equation}
\begin{split}
\Jf &= \Jf_{\rm sph}\Jf_{\rm geo}\Jf_{\rm kin},\\
\Df &= \Df_{\rm sph}\Df_{\rm geo}\Df_{\rm kin}.
\end{split}
\end{equation}

If an oblate model is viewed along the short axis, or a prolate model
is viewed along the long axis, then it appears round. The line of
sight coincides with the symmetry or $z$ axis. The geometric
corrections are then straightforward to evaluate as
\begin{equation}
 \Jf_{\rm geo,face} = {1 \over q},\quad\Df_{\rm geo,face} = {1}.
 \label{eq:Joblateshort}
 \end{equation}
This case is very simple because both the field of view and the
surface density contours are circular. Note that both factors are independent of the slope of the density profile $\gamma_\mathrm{DM}$.

If an infinite ($r_t\rightarrow\infty$) oblate or prolate model is viewed edge-on, it appears flattened
with axis ratio $q$. The line of sight then coincides with, say, the
$y$ direction. Observationally, the effective radius of a flattened model is
always measured along the projected major axis.  For an oblate model, the measured effective radius is $\rh$ whereas for the prolate model, the effective radius of its spherical progenitor is actually
$\rh/q$. Additionally, comparison with the `ellipticity-corrected' spherical model gives rise to an additional factor of $\sqrt{1-\epsilon}$ in the effective radius which equals $\sqrt{q}$ for the oblate case and $\sqrt{1/q}$ for the prolate case.

The geometric corrections (i.e. the ratio of the J- and D-factors to those for a spherical model with the same $\Mh$) are now
\begin{equation}
\Jf_{\rm geo,edge} = {q^{2-\gamma_\mathrm{DM}}\over 2 \pi q^2} \int_0^{2\pi}d\theta (\cos^2\theta +
q^{-2}\sin^2\theta)^{1/2-\gamma_\mathrm{DM}},
\label{eq:Joblatelong2}
\end{equation}
and
\begin{equation}
\Df_{\rm geo,edge} = {q^{1-\gamma_\mathrm{DM}/2}\over 2 \pi q} \int_0^{2\pi}d\theta (\cos^2\theta +
q^{-2}\sin^2\theta)^{1/2-\gamma_\mathrm{DM}/2}.
\end{equation}
The factors can only be reduced to a single quadrature due to the mismatch between the circular beam aperture and the elliptical isophotes. For oblate (prolate) models, the geometric correction leads to
an increase (decrease) in the J-factor as compared to a spherical
model with the same $\Mh$ if $\gamma_\mathrm{DM}\leq2$. If the dark matter halo is truncated at a finite ellipsoidal radius $r_t<D\theta$, the beam encloses all the dark matter and the edge-on geometric factors reduce to
\begin{equation}
 \Jf_{\rm geo,edge} = q^{1-\gamma_\mathrm{DM}},\quad\Df_{\rm geo,edge} = q^{1-\gamma_\mathrm{DM}/2}.
 \label{eq:Jedgetrunc}
 \end{equation}
These equations are preferable as for $\gamma_\mathrm{DM}<3$ they correspond to finite mass models and for $\gamma_\mathrm{DM}<3/2$ they produce finite J-factors. We have found that they give much better representations of the correction factors for more general models.

We have computed the ratio of the J- and D-factors to those of the spherical model with the same $\Mh$. As $\Mh$ is estimated from the line-of-sight velocity dispersion, we must now compute the ratio of the true $\Mh$ to that computed using only the line-of-sight velocity dispersion. This ratio is the kinematic correction, which we compute using the tensor virial
theorem~\citep[e.g.,][]{Bi78,Ag12}.  The effect of flattening on the
kinematics of the stars is given by
\begin{equation}
{T_{RR} \over T_{zz} }
= {W_{RR} \over W_{zz} }
=  {\ds\int\mathrm{d}^3\boldsymbol{x}\,\rho_\star(\sigma^{2}_{RR} + \sigma^2_{\phi\phi})
  \over \ds \int\mathrm{d}^3\boldsymbol{x}\,{\rho_\star
    \sigma^{2}_{zz}}}\ .
\label{eq:virial}
\end{equation}
where $T$ and $W$ are the kinetic energy and potential energy
tensors~\citep{Ch69,BT,Ag12}.  The stellar density in dSphs is well
approximated by a Plummer or King profile. Such laws do not lead to
tractable integrals in the virial theorem~(\ref{eq:virial}). Instead,
we approximate the stellar density as a power-law stratified on
similar concentric spheroids with $m_\star^2 = R^2 + z^2 q_\star^{-2}$
and so $q_\star$ is the stellar flattening. This means we can take
advantage of equations (19-24) in \citep{Ag12}, which give the virial
ratios for stellar populations whose density is a pure scale-free
power-law declining like distance$^{-\gamma_\star}$. Note that as the all the considered models have infinite mass we must work with the ratios of the velocity dispersions.

\begin{figure}
$$\includegraphics[width=\columnwidth]{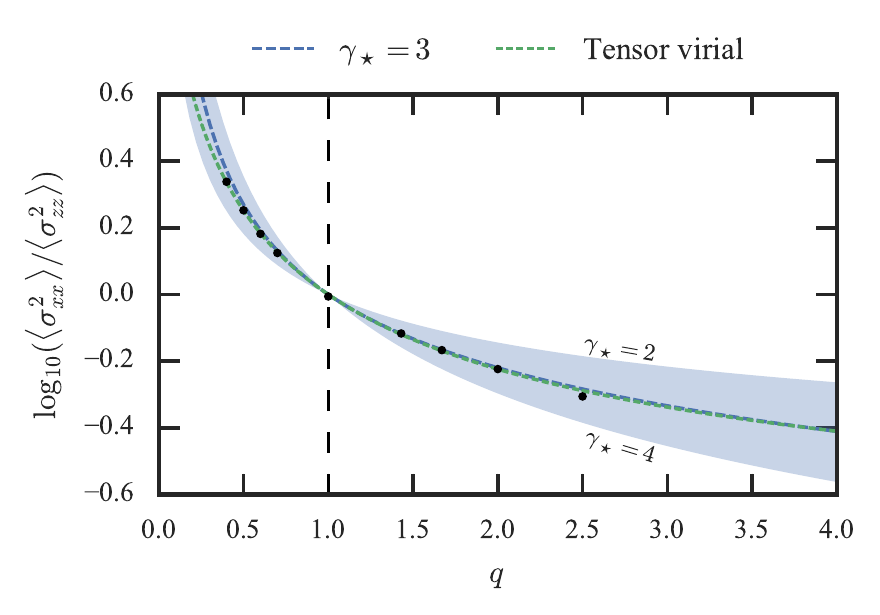}$$
\caption{Kinematic ratio for oblate and prolate figures. Each line
  shows the prediction from a stellar axisymmetric cusp with density
  slope $\gamma_\star$ flattened with axis ratio $q$ embedded in a
  halo also with flattening $q$. The black points show the numerical
  results from the M2M models.}
\label{fig:WR_corrections}
\end{figure}

If we assume the equipotentials are spheroidally stratified, the correction is a function of $Q_\star = q_\phi^2/q_\star^2$, where
$q_\phi$ is the flattening of the dark halo equipotentials, which is
related to the flattening $q$ in the dark halo density via
\begin{equation}
q_\phi = {1\over 2}\left( 1 + \left( 1 + 8q^2 \right)^{1/2}
\right)^{1/2}
\label{eq:BowdenFormula}
\end{equation}
This formula is given in refs~\citep{Ev93,Bo15}. As is well known, the
equipotentials are always rounder than the density contours, so that
$q_\phi \approx 1$ even if the dark halo is quite flattened.  Then for
$\gamma_\star =3$, we have from \citep{Ag12}
\begin{equation}
    {\langle\sigma^2_{xx}\rangle \over \langle\sigma^2_{zz}\rangle}=
    \begin{cases}
    {\displaystyle \frac{Q_\star \mathcal{Q}-\sqrt{Q_\star}\arcsinh{\mathcal{Q}}}
    {2[\sqrt{Q_\star}\arcsinh{\mathcal{Q}}-\mathcal{Q}]}},
     & Q_\star >1\\
\null & \null \cr
    {\displaystyle
    \frac{Q_\star\mathcal{Q}-\sqrt{Q_\star}\arcsin{\mathcal{Q}}}
    {2[\sqrt{Q_\star}\arcsin{\mathcal{Q}}-\mathcal{Q}]}},
     & Q_\star <1,
\end{cases}
\end{equation}
where
\begin{equation}
    \mathcal{Q}(Q_\star)=
    \begin{cases}
    \sqrt{Q_\star-1}& Q_\star>1\\
    \sqrt{1-Q_\star}& Q_\star<1\\
    \end{cases}
\label{curlyQ}
\end{equation}
Our formulae are appropriate if $r_t\rightarrow\infty$ or the stellar profile is truncated at a smaller radius than the dark-matter profile.
This virial ratio is unity when $Q_\star =1$. This follows because if
the stellar density is constant on the equipotentials, then the
velocity dispersion is isotropic. It is greater than unity when
$Q_\star >1$ (that is, when the model is oblate), as the globally
averaged velocity dispersion component along the long or $x$ axis must
be larger than that along the short or $z$ axis. It is less than unity
when $Q_\star <1$, as the roles of the $x$ and $z$ axes are now
reversed for the prolate figure.  Formulae for other values of
$\gamma_\star$, or stellar density fall-off, are given in
Appendix~\ref{CuspAppendix}.  We plot the logarithm of the virial ratio in
Fig.~\ref{fig:WR_corrections} along with the ratio calculated from the
M2M models. The line for $\gamma_\star=3$ agrees well with the M2M
data. The green dashed line shows the virial ratio computed using the virial method. For a stellar density stratified on the same concentric ellipsoids as the dark-matter density, the virial ratio is simply a function of the shape of the ellipsoids and is independent of the radial density profile so the plotted line has a very simple functional form \citep{Roberts1962,White1989,BT}.

It can be shown that along the sequence of models the total luminosity-averaged square velocity dispersion $\sigma_\mathrm{tot}$ is constant. Therefore, for a given model the kinematic factor is the ratio of the total velocity dispersion to the line-of-sight velocity dispersion. When viewed down the $x$-axis, the kinematic correction factor is
\begin{equation}
    J_{\rm kin, edge} = \Big({\langle\sigma_{\mathrm{tot}}^2 \rangle\over \langle\sigma_{\mathrm{los}}^2\rangle}\Big)^2=\Big({2\over 3} + {\langle\sigma_{zz}^2 \rangle\over 3\langle\sigma_{xx}^2\rangle}\Big)^2
\end{equation}
This is the ratio of the squared velocity dispersion along the line of
sight to the average value. This is smaller (larger) than unity for
oblate (prolate) models.  When viewed down the $z$-axis, the kinematic
correction factor is
\begin{equation}
    J_{\rm kin,face} = \Big({1\over 3} + {2\langle\sigma_{xx}^2\rangle \over 3\langle\sigma_{zz}^2\rangle}\Big)^2
\end{equation}
This is smaller (larger) than unity for prolate (oblate) models. Note that, as the D-factors are proportional to $\sigma_\mathrm{tot}^2$, the D-factor kinematic factor $D_\mathrm{kin}=\sqrt{J_\mathrm{kin}}$.

\subsection{Flat Rotation Curve Models}\label{CoredModels}

A simple but widely-used model of a dark halo has potential-density pair \citep{BT,Ev93}
\begin{eqnarray}
\rhoDM(R,z) &=& {v_0^2 \over 4 \pi G q_\phi^2}{ (2q_\phi^2 +1)\Rd^2 +
  R^2 + z^2(2-q_\phi^{-2}) \over (\Rd^2 + R^2 + z^2 q_\phi^2)^2},\nonumber\\
\Phi_\mathrm{DM}(R,z) &=& {v_0^2 \over 2} \ln (\Rd^2 + R^2 + z^2 q_\phi^{-2})
\end{eqnarray}
Here, $v_0$ is a velocity scale that is the asymptotic value of the
flat rotation curve, whilst $\Rd$ is the dark matter length-scale
while $q_\phi$ is the axis ratio of the equipotentials.  The dark
matter density is everywhere positive provided $q_\phi > 1/\sqrt{2}$,
so the model can be oblate, spherical or prolate. Unless $q_\phi=1$, the flattening of the dark matter density changes with radius such that the oblate models become more oblate in the outskirts whilst the prolate models become more prolate. At large radii $q_\phi$ is related to the isodensity flattening $q$ via equation~\eqref{eq:BowdenFormula}. The dark halo is
cusped if $\Rd=0$, but the cusp is isothermal and so much more severe
than in the NFW model.

The J-factor for the model viewed along the $z$-axis or symmetry axis is
\begin{eqnarray}
\Jf = {v_0^4\over 96 \Rd D^2 G^2 q_\phi^3}&\Bigl[& 3(1-y)
  -4q_\phi^2(y^3-1)\nonumber \\ &+&
q_\phi^4(8-3y-2y^3-3y^5)\Bigr],
\end{eqnarray}
with $y = R_d/\sqrt{R_d^2+D^2\theta^2}$. At large angles,
$y\rightarrow 0$ and so the asymptotic value is
\begin{equation}
\Jf \rightarrow  {v_0^4\over 96 \Rd D^2 G^2 q_\phi^3}[ 3
  +4q_\phi^2 +
8q_\phi^4].
\end{equation}
Similarly, the D-factor for a model viewed along the $z$-axis is given by
\begin{equation}
\Df = \frac{v_0^2 R_d}{GD^2}\frac{ q_\phi(D\theta/R_d)^2}{\sqrt{1+(D\theta/R_d)^2}}.
\end{equation}
Note that the total mass of the model is not finite so the D-factor does not tend to a finite value as $\theta\rightarrow\infty$.
Viewed edge-on, two of the integrations for the J-factor are
analytic, leaving a final integral over the spherical aperture to be
performed numerically
\begin{equation}
\begin{split}
J(\theta&\rightarrow\infty) = {v_0^4 \over 1536 \pi \Rd D^2 G^2 q_\phi^8} \\&\times\int_0^{2\pi}d\phi{
  G_1(q_\phi)+G_2(q_\phi)\cos(2\phi)+G_3(q_\phi)\cos(4\phi)
\over
  (\cos^2\phi + q_\phi^{-2} \sin^2\phi)^3},
\end{split}
\end{equation}

\begin{equation}
\begin{split}
G_1(q_\phi)&=120 + 280 \mathcal{Q}^2 + 221 \mathcal{Q}^4 + 64 \mathcal{Q}^6 + 9 \mathcal{Q}^8,\\
G_2(q_\phi)&=4 \mathcal{Q}^2 (14 + 3 \mathcal{Q}^2 (3 + \mathcal{Q}^2)^2),\\
G_3(q_\phi)&=\mathcal{Q}^4 (7 + 8 \mathcal{Q}^2 + 3 \mathcal{Q}^4),
\end{split}
\end{equation}
and $\mathcal{Q}^2=q_\phi^2-1$.
The D-factor can also be expressed as a single quadrature but the expression is too bulky to present here.
Into this dark halo, we embed a population of stars to model the dSph, namely
\[
\rho(R,z) = {\rho_0 \Rc^{\beta_\star} \over (\Rc^2 + R^2 + z^2q_\star^{-2})^{\beta_\star/2}}.
\]
Here, $\rho_0$ is a normalization constant, while $q$ is the axis
ratio of the spheroidal isodensity contours. If $\beta_\star=5$, this is the
familiar Plummer model.  If the scale-length of the stars $\Rc$ is equal
to the scale-length of the dark matter $\Rd$, and the flattening of the
stellar density $q_\star$ is equal to the flattening of the dark matter
equipotentials $q_\phi$, then the phase space distribution function
is an isothermal~\citep{Ev93}. We derive more general formulae below,
but note that this simple limit enables an easy check of the
correctness of our results.

As both the density and the potential are simple, we can calculate the
velocity dispersions seen on viewing the stellar distribution along
the short or long axis. We give the results for $\beta_\star =5$ here,
and delegate other formulae to the Appendix B. We begin by defining
\begin{equation}
\Delta_1^2 = q_\phi^2 -q_\star^2,\quad \Delta_2^2 = \Rd^2-\Rc^2,\quad
\mathcal{D}^2 = q_\phi^2\Rd^2 - q_\star^2\Rc^2.
\end{equation}
Then the velocity dispersions are
\begin{eqnarray}
\langle\sigma^2_{RR}\rangle &=&  {v_0^2 q_\phi \Rc^2 \over
  \Delta_1^4\Delta_2^4\mathcal{D}^2} \Bigl[
2 \Delta_1^4 \Rd^3 \mathcal{D}\arccosh \left({q_\phi\Rd\over
  q_\star\Rc}\right) \nonumber \\
&-&\Delta_1 \mathcal{D}^2(2q_\phi^2\Rd^2 + q_\star^2(\Rc^2\!-\!3\Rd^2))
\arccosh
\left({q_\phi\over q_\star}\right) \nonumber\\
&-& q_\phi \Delta_1^2\Delta_2^2\mathcal{D}^2\Bigr],\nonumber\\
\langle\sigma^2_{zz}\rangle &=&
{v_0^2 q_\star^2 \Rc^2 \over \Delta_1^4 \Delta_2^2 \mathcal{D}^2} \Bigl[
q_\star^2\Delta_1^2 \Delta_2^2 \mathcal{D}^2 - q_\phi\Delta_1 \mathcal{D}^4
\arccosh
\left({q_\phi\over q_\star}\right) \nonumber \\
&+& q_\phi\Delta_1^4 \mathcal{D} \Rd^3 \arccosh \left({q_\phi\Rd\over q_\star\Rc}\right)\Bigr].
\end{eqnarray}
The formulae hold generally on using the identity (for $S<1$)
\[
\arccosh S \equiv -i \arccos S
\]
These formulae give the line of sight velocity dispersion of an
axisymmetric Plummer model viewed along the short and long axes in a
dark halo of arbitrary flattening and length-scale.
If $\Rc = \Rd$, then
\begin{eqnarray}
\langle\sigma^2_{RR}\rangle &=& {v_0^2q_\phi \over 4(q_\phi^2-q_\star^2)^3}\Bigl[ 5q_\star^4q_\phi -7q_\star^2q_\phi^3
+ 2q_\phi^5 \nonumber\\
&+& 3q_\star^4(q_\phi^2-q_\star^2)^{1/2}\arccosh\left({q_\phi\over q}\right)\Bigr],
\nonumber \\
\langle\sigma^2_{zz}\rangle &=& {v_0^2q_\star^2 \over
  2(q_\phi^2-q_\star^2)^3}\Bigl[q_\phi^4 + q_\phi^2q_\star^2 - 2q_\star^4 \nonumber\\
&-& 3 q_\star^2q_\phi(q_\phi^2-q_\star^2)^{1/2}\arccosh\left({q_\phi\over q}\right)\Bigr].
\end{eqnarray}
If additionally $q_\star = q_\phi$, then
\begin{equation}
\langle\sigma^2_{RR}\rangle = {2v_0^2 \over 5},\qquad\qquad
\langle\sigma^2_{zz}\rangle = {v_0^2\over 5}.
\end{equation}

With the line of sight velocity dispersion in hand, we can simply
rescale the model so that the J-factors are computed for models with
the same observables (line of sight velocity dispersion and half-light
radius) as the flattenings and the ratio of dark to luminous
scalelength $\Rc/\Rd$ varies.

\subsection{Comparisons}

\begin{figure}!
$$\includegraphics[width=0.5\textwidth]{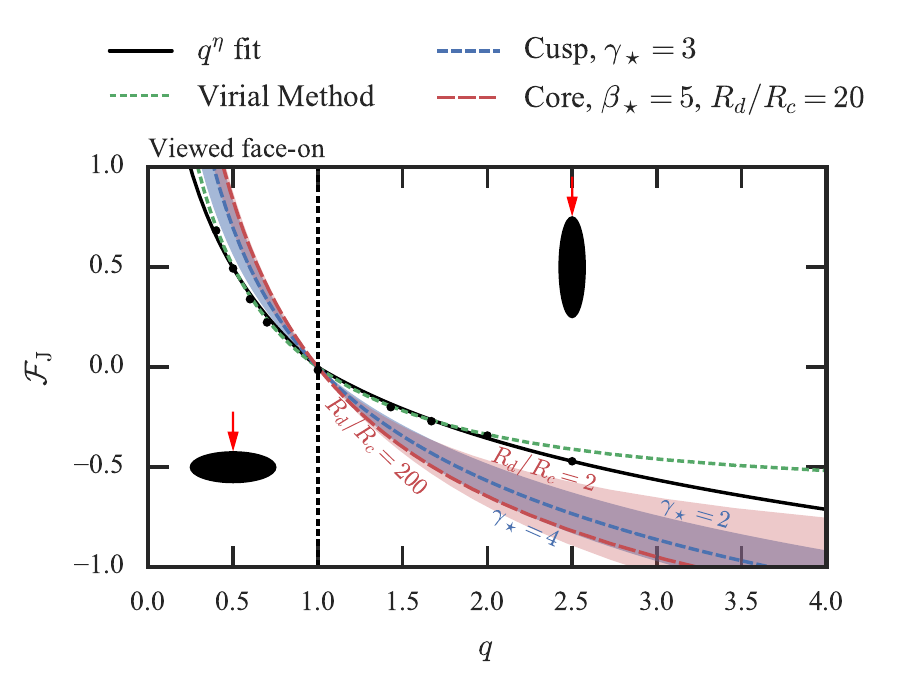}$$
$$\includegraphics[width=0.5\textwidth]{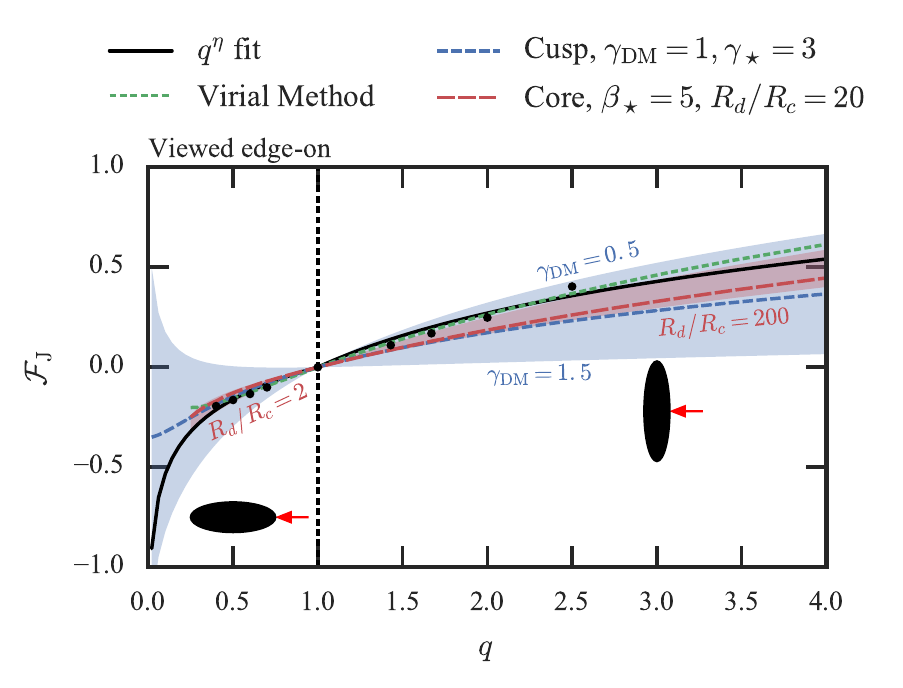}$$
\caption{J-factor correction factors for oblate and prolate figures
  viewed face-on (top) and edge-on (bottom). The \textbf{black points}
  show the numerical results from the M2M models of stellar Plummer
  models flattened with axis ratio $q$ embedded in NFW dark-matter
  halos also of axis ratio $q$. The \textbf{dashed green line} shows the results
  of the virial method of Section~\ref{VirMethod}. The \textbf{blue band} shows a range of axisymmetric cusp models from Section~\ref{AxiCusps}. The central line corresponds to a model with $\gamma_\mathrm{DM}=1,\,\gamma_\star=3$. In the top panel, we have varied the slope of the light profile (note the face-on correction factor is independent of $\gamma_\mathrm{DM}$ in this case). In the bottom panel, we have varied the slope of the dark matter. In both panels, the \textbf{red band} shows a series of cored flat rotation curve models from Section~\ref{CoredModels}. The central line has outer stellar density profile of $\beta_\star=5$ and ratio of dark-matter to stellar scale radii of $R_\mathrm{d}/R_\mathrm{c}=20$. The band corresponds to varying this scale radii ratio by a factor of ten.}
\label{fig:corrections}
\end{figure}

\begin{figure}!
$$\includegraphics[width=0.5\textwidth]{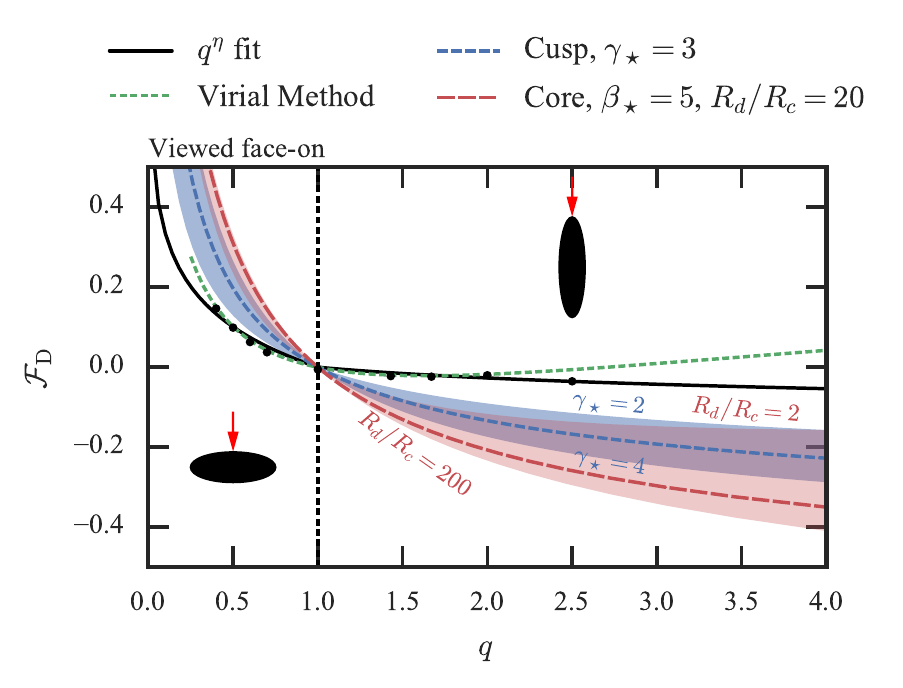}$$
$$\includegraphics[width=0.5\textwidth]{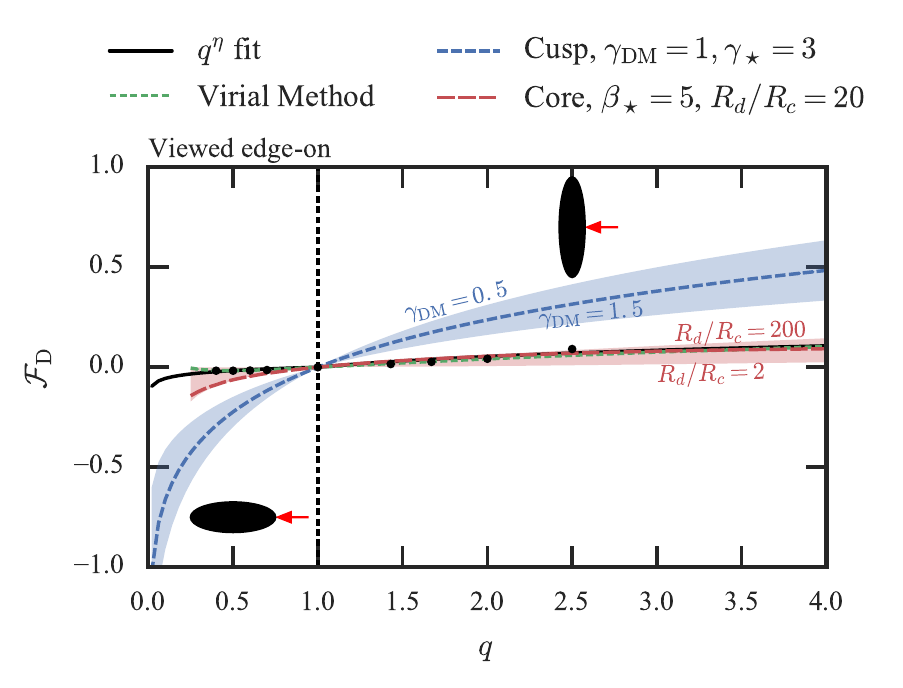}$$
\caption{D-factor correction factors for oblate and prolate figures
  viewed face-on (top) and edge-on (bottom). See the caption of Figure~\ref{fig:corrections} for details on each line.}
\label{fig:D_corrections}
\end{figure}

The models in Sections~\ref{AxiCusps} and \ref{CoredModels} are
complementary. The infinite cusps allow us to vary the central slope of
the dark matter. The cored models allow us to vary the ratio of the
luminous to the dark matter length-scale. Taken together, a gamut of
possibilities of dark halo cusps, density profiles and length-scales
can be swept out.

The base-10 logarithms of the correction factors~(\ref{eq:defcorrfactors}) are plotted as a
function of dark halo flattening $q$ in
Fig.~\ref{fig:corrections}. The flattening of the stellar population
$q_\star$ is the same as that of the dark halo $q$. For all plots we use the observed parameters of Reticulum~II with a beam angle of $0.5^\circ$. The correction factors are a function of the beam angle but we have found that this dependence is very weak and that the correction factors are essentially independent of the size of the dSph with respect to the beam size. Therefore, the reported correction factors are appropriate for all beam angles.

We show the
correction factors for models viewed face-on (top panels) and edge-on
(bottom panels). Any correction factor between these extremes should
be possible as the inclination angle can be varied between these
extremes. The correction factors computed from the M2M Plummer models
in NFW halos with axis ratio $q$ are shown as black points. They are
in good agreement with the results of the virial method of
Section~\ref{VirMethod} applied to the self-same model, which are
shown as a green curve. Note that the exception to this is the $q=2.5$ prolate model. As noted in Section~\ref{Sect::M2M}, the Made-to-Measure method in this case produces an equilibrium figure that significantly deviates from a spheroidally stratified Plummer model. We also show as blue bands the range of
results for the axisymmetric cusps of Section~\ref{AxiCusps}, in which
both the slopes of the dark matter and stellar cusps are allowed to
vary. Finally, the red band shows a series of cored flat rotation
curve models from Section~\ref{CoredModels}. The central line has
outer stellar density slope of $\beta_\star=5$ and ratio of
dark-matter to stellar scale radii of
$R_\mathrm{d}/R_\mathrm{c}=20$. The band corresponds to varying this
scale radii ratio by a factor of ten. As we move along the curves, the
models have the same line-of-sight velocity dispersion and the same
half-light radius.  As the models make different assumptions as to the
dark matter density and potential, we do not expect these curves to
match up exactly with the M2M models, but it is encouraging that they
all show similar trends.

When an oblate model is viewed along the short axis or face-on, it
appears circular, but there is always a boost to the J-factor. For
flattenings of $q = q_\star = 0.5$, this can be a factor of $3$ boost over the spherical J-factor.  When the model is viewed
edge-on, or along the long axis, then it appears flattened with
isophotes of ellipticity $1-q_\star$.  However, the geometric and
kinematic corrections work in different directions, the former to
boost, the latter to reduce, the J-factor. The net effect is less significant than in the face-on case and is $\sim0.1-0.2\dex$ for $0.4<q<0.7$.

When a prolate model is viewed along the long or $z$-axis, it appears
round. Here, the geometric and the kinematic corrections both diminish
the J-factor. Although the model looks round on the sky, its J-factor
can be substantially less than computed by a spherical analysis. For
example, if the true flattening is $q = q_\star = 2.5$, then the
J-factor is decreased by a factor $\lesssim 0.3$. When the model is
viewed edge-on, the isophotes have ellipticity $1-q_\star^{-1}$, and the kinematic and geometric factors act in opposite directions with the net result being a small boost. If the flattening is $q = q_\star = 2.5$ the J-factor is increased by $\sim 0.3\dex$ over the spherical estimate.

The blue bands give an indication of how the correction factors vary as the dark-matter density slope is adjusted. We find steeper cusps give smaller corrections for models viewed edge-on, but has no effect on the corrections for models viewed face-on as the geometric factor is independent of $\gamma_\mathrm{DM}$. For the face-on case we see that making the slope of the stellar density profile steeper produces larger corrections to the J-factor.

From the red bands we observe that making the dark matter halo more extended (increasing $\rd/\Rc$) produces larger corrections for the face-on case but smaller corrections for the edge-on case. The width of the red bands when the length-scales $\Rc/\Rd$ is varied
is at most 0.5 dex, even at the most extreme flattenings. Most of the
Milky Way dSphs are rounder than $q_\star = 0.5$. In this regime, the
red band is thinner, and gives rise to an uncertainty of at most $\sim
0.25$ dex. This suggests that varying the concentration of the dark
matter halo will not have a significant effect on the flattened
J-factors. This is corroborated by experiments with the Plummer profile embedded in the NFW profile. Additionally, making the stellar density in these models fall off more rapidly increases the magnitude of the J-factor correction factors when viewing face-on but decreases the magnitude of the correction when viewing edge-on.

The equivalent results for the D-factors are shown in
Fig.~\ref{fig:D_corrections}. The correction factors for the M2M models
models are all $\lesssim0.2\dex$, and suggest that for most applications, the spherical
approximation suffices for the D-factors. Note that in the face-on case, the cored models give a similar approximation of the correction factors as the cuspy models whilst for the edge-on case the cored models more faithfully represent the true correction factors than the cuspy models. Increasing the outer stellar density slope for the flat rotation curve models increases the magnitude of D-factor correction factors when viewing face-on but has little effect for the edge-on case.

Finally, we note that for an $(\alpha,\beta,\gamma)$ stellar model of equation~\eqref{abgmodel} embedded in another $(\alpha,\beta,\gamma)$ dark-matter model the J-factor correction factors are very insensitive to the choice of the density slopes of the stellar and dark-matter distributions and the ratio of the stellar to dark-matter scale-lengths. The same is broadly true for the D-factor correction factors except that the edge-on correction factor has a weak dependence with the outer slope of the dark matter profile. This is slightly at odds with the flat rotation curve model results but this may be due to the flat rotation curve models having a density flattening that varies with radius whilst the $(\alpha,\beta,\gamma)$ models have a constant density flattening.

\section{The Effects of Triaxiality}

Generically, we might expect the light and dark-matter distributions in dwarf spheroidals to be triaxial \cite{SanchezJanssen2016}. Triaxiality can introduce additional flattening (stretching) along the line-of-sight and so naturally increases (decreases) the J-factor and gives rise to larger (smaller) correction factors. Here we extend the formulae given in the previous section to account for intrinsic triaxial shapes. We begin by focusing on the infinite cusp models where some analytic progress can be made before moving on to consider more general density profiles.

We extend the models of equation~\eqref{AxiCusp} and introduce an intermediate-to-major axis ratio $p$ in addition to the minor-to-major axis ratio $q$. Here we restrict $q<p<1$ such that a prolate model has $p=q\neq1$. It is conventional to use a triaxiality parameter $T$ to describe the figures defined by
\begin{equation}
T = \frac{1-p^2}{1-q^2}.
\end{equation}
Note that figures with $T=0$ are oblate whilst those with $T=1$ are prolate. The density for the triaxial cusp models is

\begin{equation}
\rhoDM(m) = {\Mh\over 4 \pi pq\mh^{3-\gamma_\mathrm{DM}}}{3 -\gamma_\mathrm{DM} \over m^{\gamma_\mathrm{DM}}}\quad \mathrm{for}\quad m\leq r_t,
\label{TriaxCusp}
\end{equation}
and zero otherwise. $m^2 = x^2 + y^2p^{-2} + z^2q^{-2}$ and $r_t$ is a truncation ellipsoidal radius.
When an infinite ($r_t\rightarrow\infty$) model is viewed along the $z$ axis, the observed flattening is $p$ and the geometric factor is a combination of equation~\eqref{eq:Joblateshort} and~\eqref{eq:Joblatelong2} such that
\begin{equation}
\Jf_{\rm geo,z} = {p^{2-\gamma_\mathrm{DM}}\over 2 \pi p^2 q} \int_0^{2\pi}d\theta (\cos^2\theta +
p^{-2}\sin^2\theta)^{1/2-\gamma_\mathrm{DM}}.
\end{equation}
In this case, the observed major-axis length corresponds to the intrinsic model scale radius. For $\gamma_\mathrm{DM}\leq2$, the integral is a monotonic function of $q$ that is greater than unity for $q<1$ and less than unity for $q>1$. If viewed along the $y$-axis the observed flattening is $q$ and the J-factor is given by

\begin{equation}
\Jf_{\rm geo,y} = {q^{2-\gamma_\mathrm{DM}}\over 2 \pi q^2 p} \int_0^{2\pi}d\theta (\cos^2\theta +
q^{-2}\sin^2\theta)^{1/2-\gamma_\mathrm{DM}},
\end{equation}
and again the observed major-axis length coincides with the intrinsic model scale radius. When viewed along the major-axis, the observed flattening is $q/p$ and the observed major-axis length coincides with the intermediate axis so the resultant measured scale-length must be scaled by a factor $1/p$. This gives rise to a geometric factor of
\begin{equation}
\Jf_{\rm geo,x} = {(q/p)^{2-\gamma_\mathrm{DM}}\over 2 \pi (q/p)^2 p} \int_0^{2\pi}d\theta (\cos^2\theta +
(q/p)^{-2}\sin^2\theta)^{1/2-\gamma_\mathrm{DM}}.
\end{equation}
As with the axisymmetric case, the infinite cusps have limited use and it is more practical to use models with finite truncation ellipsoidal radii $r_t<D\theta$. In this case, the geometric factors are given by
\begin{equation}
\begin{split}
\Jf_{\rm geo,x}&=(qp)^{1-\gamma_\mathrm{DM}},\\
\Jf_{\rm geo,y}&=q^{1-\gamma_\mathrm{DM}}/p, \\
\Jf_{\rm geo,z}&=p^{1-\gamma_\mathrm{DM}}/q.
\end{split}
\end{equation}
For the models that produce a finite J-factor ($\gamma_\mathrm{DM}<3/2$), we find $\Jf_{\rm geo,x}<\Jf_{\rm geo,y}<\Jf_{\rm geo,z}$. For the astrophysically-motivated case of $\gamma_\mathrm{DM}=1$ the geometric factors are simply $\Jf_{\rm geo,x}=1$, $\Jf_{\rm geo,y}=1/p$ and $\Jf_{\rm geo,z}=1/q$.

If the infinite model is observed along a line of sight oriented with spherical polar angles $(\varphi,\vartheta)$ with respect to the intrinsic Cartesian coordinates of the model, the geometric factor must be computed with the full three-dimensional integrals as
\begin{equation}
\Jf_{\mathrm{geo}} = \frac{1}{\Jf_\mathrm{sph} D^2}\int_{-\infty}^{\infty}\mathrm{d}z'\,\int_{0}^{2\pi}\mathrm{d}\theta\,\int_{0}^{D\alpha}\mathrm{d}R\,R\rhoDM^2(\boldsymbol{x}),
\end{equation}
where
$
 \boldsymbol{x} = R\cos\theta \hat{\boldsymbol{\varphi}}+R\sin\theta\hat{\boldsymbol{\vartheta}}+z'\hat{\boldsymbol{r}}
$
with $(\hat{\boldsymbol{r}},\hat{\boldsymbol{\varphi}},\hat{\boldsymbol{\vartheta}})$ the set of spherical polar unit vectors. Making the model finite with $r_t<D\theta$ produces $\Jf/\Jf_{\rm sph}=1/(pq)$. However, for this general viewing angle calculation of the observed scale radius seems intractable. The kinematic factors $\Jf_\mathrm{kin} = (\langle\sigma_\mathrm{tot}^2\rangle/\langle\sigmalos^2\rangle)^2$ can be derived for these more general viewing angles as
\begin{equation}
\Jf_\mathrm{kin} = \Big[\frac{1}{3}\frac{1+f_1+f_2}{\cos^2\vartheta+f_1\sin^2\vartheta\cos^2\varphi +f_2\sin^2\vartheta\sin^2\varphi }\Big]^2,
\end{equation}
where
\begin{equation}
\begin{split}
f_1&=\frac{\langle\sigma_{xx}^2\rangle}{\langle\sigma_{zz}^2\rangle} =\frac{\int_0^\pi\mathrm{d}\theta\,\int_0^{2\pi}\mathrm{d}\phi\,F(\theta,\phi)\sin^3\theta\cos^2\phi }{\int_0^\pi\mathrm{d}\theta\,\int_0^{2\pi}\mathrm{d}\phi\,F(\theta,\phi)\sin\theta\cos^2\theta }>f_2\\
f_2&=\frac{\langle\sigma_{yy}^2\rangle}{\langle\sigma_{zz}^2\rangle} =\frac{\int_0^\pi\mathrm{d}\theta\,\int_0^{2\pi}\mathrm{d}\phi\,F(\theta,\phi) \sin^3\theta\sin^2\phi}{\int_0^\pi\mathrm{d}\theta\,\int_0^{2\pi}\mathrm{d}\phi\,F(\theta,\phi) \sin\theta\cos^2\theta}>1,
\end{split}
\end{equation}
and \begin{equation}
F(\theta,\phi) = (\sin^2\theta\cos^2\phi+P_\star^2\sin^2\theta\sin^2\phi+Q_\star^2\cos^2\theta)^{-\gamma_\mathrm{DM}/2}.
\end{equation}
Here $P_\star=p_\phi/p$ and $Q_\star=q_\phi/q$ with $p_\phi$ and $q_\phi$ being the axis ratios of the dark matter potential. We see that when viewing down the major axis ($\vartheta=\pi/2,\varphi=0$) the kinematic correction factor is less than unity whilst viewing down the minor axis ($\vartheta=0,\varphi=0$) produces a kinematic correction factor greater than unity. Generally, we find that $\Jf_{\rm kin,x}<\Jf_{\rm kin,y}<\Jf_{\rm kin,z}$ such that the total correction factors for $\gamma_\mathrm{DM}<3/2$ obey the hierarchy $\mathcal{F}_\mathrm{J,x}<\mathcal{F}_\mathrm{J,y}<\mathcal{F}_\mathrm{J,z}$. We have found that the effects of triaxiality seem to be in accordance with our expectation from the axisymmetric case. When there is additional flattening along the line-of-sight the geometric and kinematic correction factors combine to increase the correction factor, whilst additional stretching decreases the correction factor.

\begin{figure}
$$\includegraphics[width=\columnwidth]{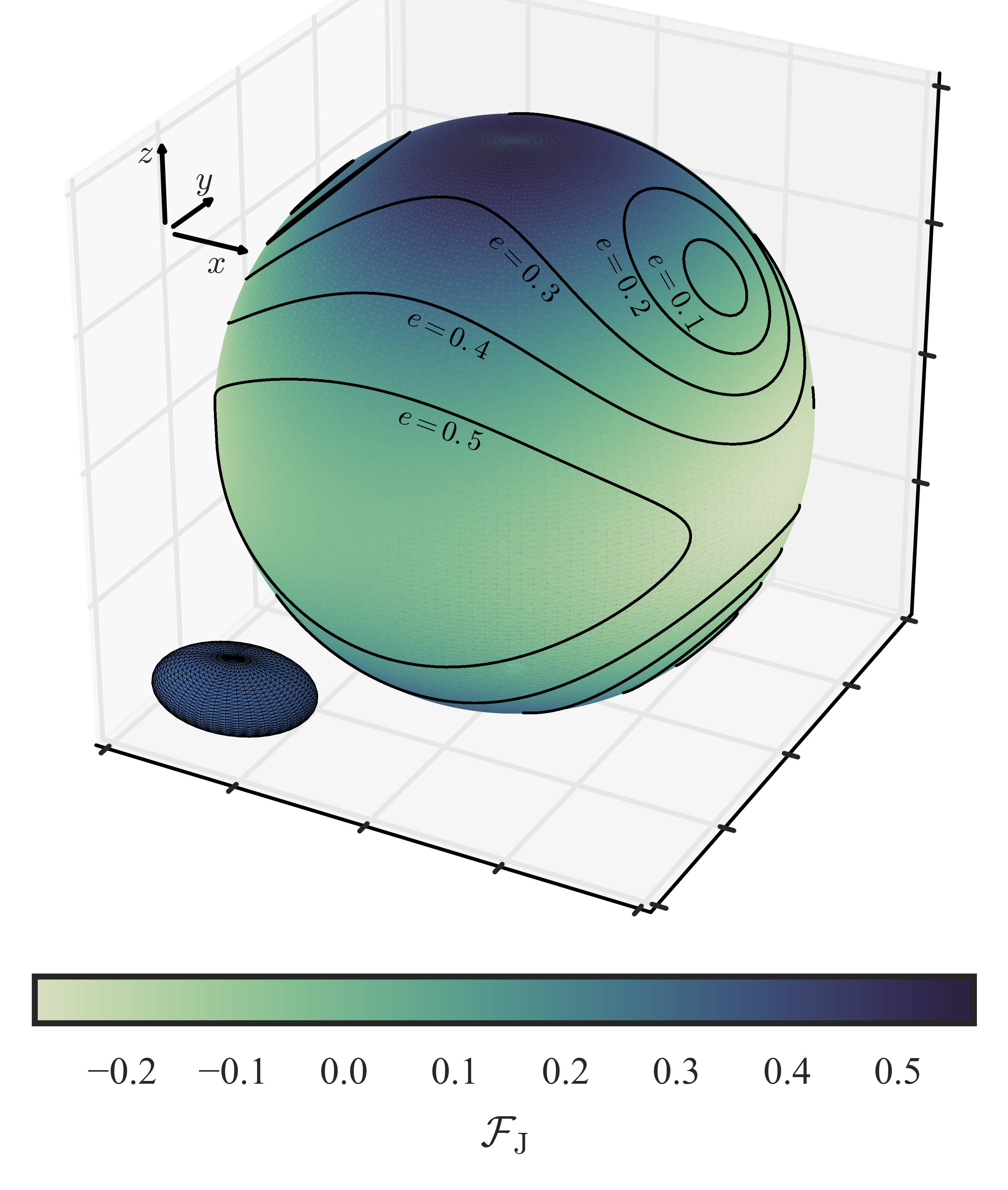}$$
\caption{J-factor correction factors for a triaxial model of Reticulum~II: each point on the sphere is colored by the correction factor when viewing the model along the radial vector that passes through that point. The black contours show the observed ellipticity when viewed from that direction. The small ellipsoid shows an isodensity contour for the considered model which has axis ratios $p=0.73$ and $q=0.4$ in both the stellar and dark-matter distributions. The largest correction factor is achieved when viewing the model down the short axis ($z$) whilst the smallest correction factor is achieved when viewing the model down the long axis ($x$). When viewing down the intermediate axis ($y$) the observed ellipticity matches that of Reticulum~II.}
\label{Triax}
\end{figure}

We now compute general triaxial correction factors using the method of Section~\ref{VirMethod}. We show an example of the J-factor correction factors for the Reticulum~II model presented in Section~\ref{Sect::M2M} but with stellar minor-to-major axis ratio $q=0.4$ and stellar intermediate-to-major axis ratio $p=0.73$. We assume the dark-matter distribution is flattened in the same way as the stellar distribution. This model has triaxiality parameter $T=0.55$, (which was deemed the best-fit to the Local Group dSphs by \cite{SanchezJanssen2016}). The base-10 logarithm of the correction factor for all viewing angles is given in Fig.~\ref{Triax}. We see that, in agreement with the simple predictions from the infinite cusp models, the largest correction factor occurs when the model is viewed down the short axis (the $z$ axis) whilst the smallest is when viewing down the long axis (the $x$ axis). The black contours on the sphere show lines of constant observed ellipticity. We see that for this figure an observed ellipticity of $0.3$ gives rise to a variation in the correction factor of $0.6\dex$.

In conclusion, additional flattening along the line-of-sight can lead to an increase in the correction factors. For a general triaxial figure the largest correction factor is obtained when viewing the model down the short axis whilst the smallest correction factor is yielded when viewing the model down the long axis.

\section{Intrinsic shapes and axis alignments of dwarf spheroidal galaxies}
We have presented corrections to the J- and D-factors based on the assumption that the dSphs are prolate or oblate figures with axes aligned with the line-of-sight. Such configurations are quite unlikely as we anticipate that generically the dSph principal axes are misaligned with the line-of-sight. In this section we will discuss what is known regarding the intrinsic shapes of the dSphs and how this translates into observed properties via their axes alignment with respect to the line-of-sight.

For a given individual galaxy, we have a couple of probes of its intrinsic shape \citep{Franx1991,Statler1994,vdB2009}. The first of these is the presence of isophotal twisting, that is the change in the orientation of the major axis of the isodensity contours with on-sky distance from the galaxy center. Isophotal twisting is a clear signature of a triaxial figure with varying axis ratios with radius, although isophotal twisting may also be caused by tidal effects \citep{Kormendy1982}. Another indicator of triaxiality is evidence of \emph{kinematic misalignment} between the axis of rotation and the minor axis of the projected density.

For entire populations of galaxies, progress can be made by analyzing statistics of the population \citep[e.g.][]{Weijmans2014}. Recently, \cite{SanchezJanssen2016} demonstrated that under the assumption that the intrinsic axes of the dSphs are randomly oriented, the Local Group dSph population is best reproduced by triaxial models with mean triaxiality $\bar T=0.55^{+0.21}_{-0.22}$ and a mean intrinsic ellipticity ($E=1-(c/a)$) of $\bar E = 0.51^{+0.07}_{-0.06}$. The assumption of random orientation is perhaps to be questioned, particularly for the Milky Way dSphs. Dark-matter only simulations \citep{Kuhlen2007,Barber2015} have demonstrated that the major axes of subhalos tend to be aligned with the radial direction to the center of their host halo and this picture has been corroborated when baryons have been included \citep{Knebe2010}. The main exception to this is near the subhalo's pericentric passage where the major axis is briefly aligned perpendicular to the radial direction.

Most of the dSphs are distant enough for the radial direction and our line-of-sight to approximately coincide, which suggests that for many dSphs the observed flattening corresponds to the intermediate-to-minor axis ratio and that there is significant stretching of the dSphs along the line-of-sight. As demonstrated in this paper, this gives rise to overestimates of the J-factors from spherical analyses for the prolate face-on models and for the triaxial model viewed down the major axis.

Based on this discussion, we now compute the expected J correction factors with their associated uncertainties under a number of assumptions regarding the intrinsic shape and alignment of the dSphs. We use the \emph{emcee} package from \cite{ForemanMackey2012} to draw $500$ samples of $(T,E,\vartheta,\varphi)$ i.e. the triaxiality, the intrinsic ellipticity and the two viewing angles. Our likelihood is the distribution of the observed ellipticity for each dSph given by, for instance, equation (A1,A2) of \cite{Weijmans2014}. For those dSphs with upper-bounds on their ellipticity we use a normal distribution with mean zero and standard deviation of half the upper-bound. We consider three different prior distributions on the parameters $(T,E,\vartheta,\varphi)$:
\begin{enumerate}
\item Uniform (U): $T\sim\mathcal{U}(0,1)$, $E\sim\mathcal{U}(0,0.95)$, $\cos\vartheta\sim\mathcal{U}(0,1)$, $\varphi\sim\mathcal{U}(0,\pi/2)$,
\item Viewing down the major-axis (R): $T\sim\mathcal{U}(0,1)$, $E\sim\mathcal{U}(0,0.95)$, $\vartheta\sim\mathcal{N}(\pi/2,0.1\,\mathrm{rad})$, $\varphi\sim\mathcal{N}(0,0.1\,\mathrm{rad})$,
\item \citet{SanchezJanssen2016} priors (T): $T\sim\mathcal{N}(0.55,0.04)$, $E\sim\mathcal{N}(0.51,0.12)$, $\cos\vartheta\sim\mathcal{U}(0,1)$, $\varphi\sim\mathcal{U}(0,\pi/2)$,
\end{enumerate}
where $\mathcal{U}(a,b)$ is a uniform distribution from $a$ to $b$ and $\mathcal{N}(\mu,\sigma)$ is a normal distribution with mean $\mu$ and standard deviation $\sigma$. For each sample we compute the base-10 logarithm of the correction factor $\mathcal{F}_\mathrm{J}$ to construct a distribution of correction factors.

\begin{figure}
$$\includegraphics[width=\columnwidth]{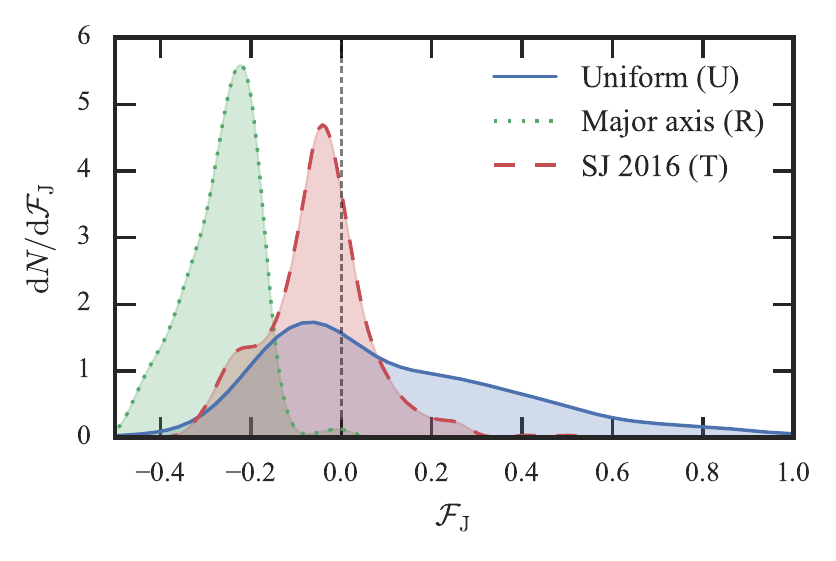}$$
\caption{Distribution of the base-10 logarithm of the J-factor correction factors for triaxial models of Reticulum~II under three different assumptions on the prior distributions of the intrinsic triaxiality, ellipticity and viewing angles as described in the text.}
\label{RetII_hr}
\end{figure}

In Figure~\ref{RetII_hr}, we show the full 1D distributions of the correction factors for Reticulum~II. All three prior assumptions produce a correction factor distribution that peaks near zero. The broadest distribution corresponds to the case where uniform priors have been adopted in all parameters. In this case, the largest correction factors correspond to models with high intrinsic ellipticity $E$ viewed down the short axis. These models have triaxiality $T$ close to unity so are near prolate models. The smallest correction factors correspond to models with low intrinsic ellipticity viewed down the long-axis.

For the prior assumption that we are viewing along the major axis, we find the median correction factor peaks at $\sim-0.2\mathrm{dex}$. For this prior assumption, there is an approximate one-to-one relationship between $T$ and $E$ as well as $T$ and $\mathcal{F}_\mathrm{J}$. Models with smaller $T$ correspond to smaller $E$ and hence smaller amplitude correction factors as these models are approximately edge-on oblate, whilst larger $T$ and larger $E$ produce larger amplitude negative correction factors as these models are nearer prolate stretched along the line-of-sight. For the prior assumption that the models have some fixed triaxiality and intrinsic ellipticity, the largest correction factors correspond to viewing angles nearer the short axis and the smallest correction factors correspond to viewing angles closer to the long axis.

\begin{figure*}
$$\includegraphics[width=\textwidth]{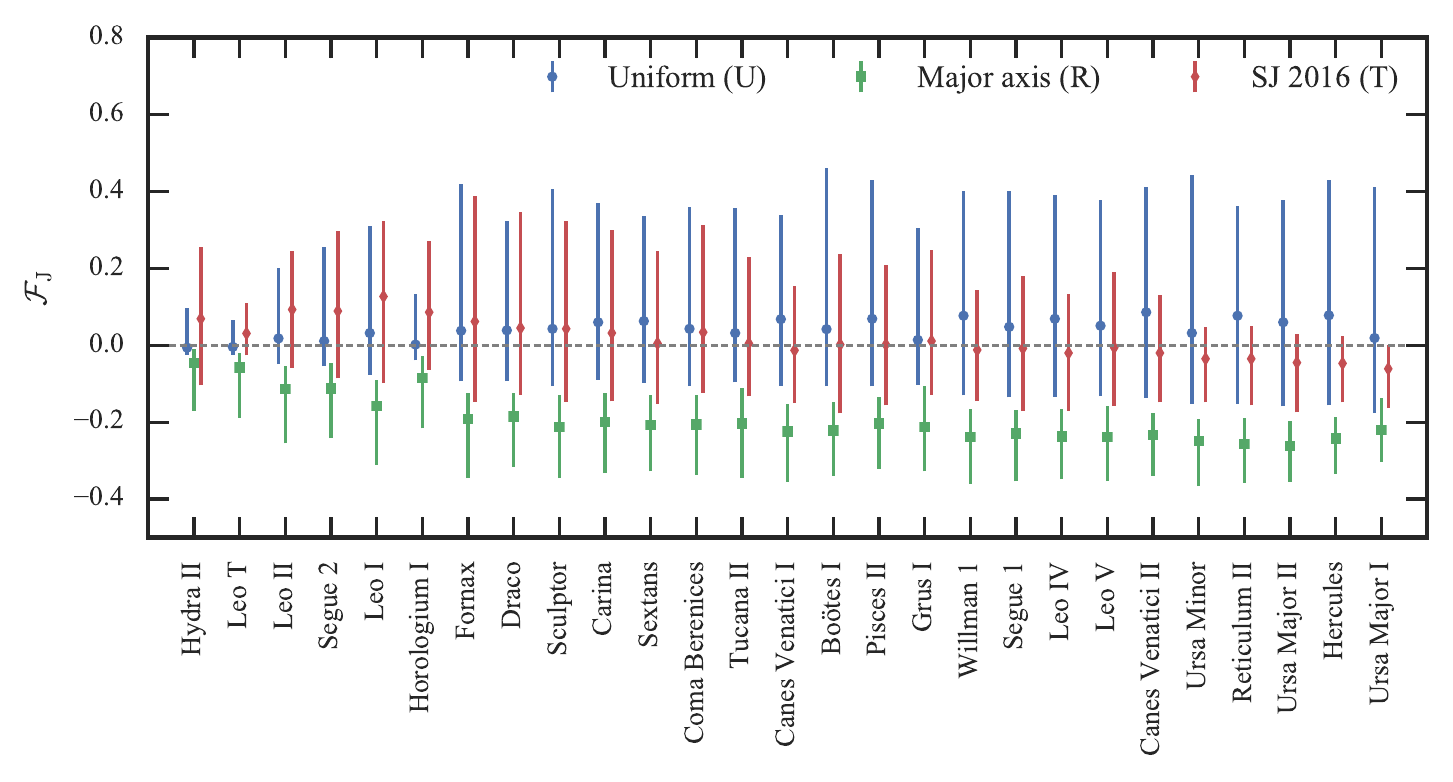}$$
\caption{Medians and $\pm1\sigma$ error-bars for the base-10 logarithms of the J-factor correction factors for all the dSphs ranked by their ellipticity. The three different error-bars correspond to three different prior assumptions regarding the intrinsic shapes and alignments of the dSphs as described in the text.}
\label{corr_triax}
\end{figure*}

The medians and $\pm1\sigma$ error-bars of the correction factors for all the dSphs computed for the three prior assumptions are given in Table~\ref{TableTriax}. This information is also displayed in Figure~\ref{corr_triax}. If we assume the dSphs are preferentially viewed along the major axis, the median correction factor is less than unity and is weakly decreasing with increasing ellipticity. As the dSphs become more flattened on the sky, they are forced to become more extended along the line-of-sight and so the J-factor decreases. The medians of the correction factors for dSphs with small ellipticity under the uniform prior assumption is around $0.05\dex$  and the upper error-bars are in general larger than the lower error-bars but the spread encompasses zero. The median shift is due to the asymmetry in the correction factors between oblate and prolate models seen in Fig.~\ref{fig:corrections}. This suggests that all J-factors are underestimated by $\sim10$ percent but naturally this conclusion is very sensitive to the exact prior assumptions. For small ellipticity, the spread in the correction factors for the uniform and fixed shape priors are approximately equal with the spread weakly increasing with increasing ellipticity for the uniform case. For the fixed shape prior the spread decreases at large ellipticity as models viewed down the minor axis become inconsistent with the observed ellipticity. We have fitted a simple relation to the uncertainty in the correction factors from the uniform priors  $\Delta \mathcal{F}_\mathrm{J}$ (the average of the $\pm1\sigma$ uncertainties) as a function of ellipticity as
\begin{equation}
\Delta\mathcal{F}_\mathrm{J} \approx 0.4\sqrt{\epsilon},
\end{equation}
which gives a fractional uncertainty in the J-factor of
\begin{equation}
\frac{\Delta \Jf}{\Jf} \approx 10^{0.4\sqrt{\epsilon}}-1.
\end{equation}
This expression gives $50\percent$ uncertainty for $\epsilon\approx0.2$, a factor $1.8$ uncertainty for $\epsilon\approx0.4$ and a factor $2.3$ uncertainty for $\epsilon\approx0.6$. For small $\epsilon$, $\Delta \Jf/\Jf\approx 0.9\epsilon$. We conclude for a typical dSph ellipticity of $0.4$ there is approximately a factor of two uncertainty in the J-factors due to the unknown triaxiality and alignment of the dSph.

\begin{table*}
\caption{Median and $\pm1\sigma$ J correction factors for the known dSphs for three different assumptions about the intrinsic shapes and alignments. The first correction factor (marked `U') uses flat uniform priors on the triaxiality, minor-to-major axis ratio and viewing angles. The second (marked `R') uses a normal prior on the viewing angles centered on $(\theta=\pi/2,\phi=0)$ with width $5^\circ$. The third (marked `T') uses a normal prior on the triaxiality and minor-to-major axis ratios with means $(0.55,0.49)$ and widths $(0.04,0.12)$ (based on the fits to the Local Group dSphs of \citet{SanchezJanssen2016}.}
\begin{tabular}{lcccc}
\hline
\hline
Name & Ellipticity & $\mathcal{F}_{\mathrm{J},U}$& $\mathcal{F}_{\mathrm{J},R}$& $\mathcal{F}_{\mathrm{J},T}$\\ 
\hline
Hydra II&$0.01^{+0.20}_{-0.01}$&$-0.01^{+0.10}_{-0.02}$&$-0.05^{+0.04}_{-0.12}$&$0.07^{+0.19}_{-0.17}$\\
Leo T&$<0.10$&$-0.00^{+0.07}_{-0.02}$&$-0.06^{+0.04}_{-0.13}$&$0.03^{+0.08}_{-0.06}$\\
Leo II&$0.13^{+0.05}_{-0.05}$&$0.02^{+0.18}_{-0.07}$&$-0.11^{+0.06}_{-0.14}$&$0.09^{+0.15}_{-0.15}$\\
Segue 2&$0.15^{+0.10}_{-0.10}$&$0.01^{+0.25}_{-0.07}$&$-0.11^{+0.07}_{-0.13}$&$0.09^{+0.21}_{-0.17}$\\
Leo I&$0.21^{+0.03}_{-0.03}$&$0.03^{+0.28}_{-0.11}$&$-0.16^{+0.07}_{-0.15}$&$0.13^{+0.20}_{-0.23}$\\
Horologium I&$<0.28$&$0.00^{+0.13}_{-0.04}$&$-0.08^{+0.06}_{-0.13}$&$0.09^{+0.19}_{-0.15}$\\
Fornax&$0.30^{+0.01}_{-0.01}$&$0.04^{+0.38}_{-0.13}$&$-0.19^{+0.07}_{-0.15}$&$0.06^{+0.33}_{-0.21}$\\
Draco&$0.31^{+0.02}_{-0.02}$&$0.04^{+0.28}_{-0.13}$&$-0.19^{+0.06}_{-0.13}$&$0.05^{+0.30}_{-0.17}$\\
Sculptor&$0.32^{+0.03}_{-0.03}$&$0.04^{+0.36}_{-0.15}$&$-0.21^{+0.08}_{-0.13}$&$0.04^{+0.28}_{-0.19}$\\
Carina&$0.33^{+0.05}_{-0.05}$&$0.06^{+0.31}_{-0.15}$&$-0.20^{+0.08}_{-0.13}$&$0.03^{+0.27}_{-0.18}$\\
Sextans&$0.35^{+0.05}_{-0.05}$&$0.06^{+0.27}_{-0.16}$&$-0.21^{+0.08}_{-0.12}$&$0.00^{+0.24}_{-0.16}$\\
Coma Berenices&$0.38^{+0.14}_{-0.14}$&$0.04^{+0.32}_{-0.15}$&$-0.21^{+0.08}_{-0.13}$&$0.03^{+0.28}_{-0.16}$\\
Bo\"otes I&$0.39^{+0.06}_{-0.06}$&$0.04^{+0.42}_{-0.15}$&$-0.22^{+0.07}_{-0.12}$&$0.00^{+0.24}_{-0.18}$\\
Canes Venatici I&$0.39^{+0.03}_{-0.03}$&$0.07^{+0.27}_{-0.17}$&$-0.22^{+0.07}_{-0.13}$&$-0.01^{+0.17}_{-0.14}$\\
Tucana II&$0.39^{+0.10}_{-0.20}$&$0.03^{+0.33}_{-0.13}$&$-0.20^{+0.09}_{-0.14}$&$0.01^{+0.22}_{-0.14}$\\
Pisces II&$0.40^{+0.10}_{-0.10}$&$0.07^{+0.36}_{-0.18}$&$-0.20^{+0.07}_{-0.12}$&$0.00^{+0.21}_{-0.16}$\\
Grus I&$0.41^{+0.20}_{-0.28}$&$0.01^{+0.29}_{-0.12}$&$-0.21^{+0.11}_{-0.12}$&$0.01^{+0.24}_{-0.14}$\\
Willman 1&$0.47^{+0.08}_{-0.08}$&$0.08^{+0.33}_{-0.21}$&$-0.24^{+0.07}_{-0.12}$&$-0.01^{+0.16}_{-0.13}$\\
Segue 1&$0.48^{+0.13}_{-0.13}$&$0.05^{+0.35}_{-0.18}$&$-0.23^{+0.06}_{-0.12}$&$-0.01^{+0.19}_{-0.16}$\\
Leo IV&$0.49^{+0.11}_{-0.11}$&$0.07^{+0.32}_{-0.20}$&$-0.24^{+0.07}_{-0.11}$&$-0.02^{+0.15}_{-0.15}$\\
Leo V&$0.50^{+0.15}_{-0.15}$&$0.05^{+0.33}_{-0.18}$&$-0.24^{+0.08}_{-0.11}$&$-0.01^{+0.20}_{-0.15}$\\
Canes Venatici II&$0.52^{+0.11}_{-0.11}$&$0.09^{+0.33}_{-0.22}$&$-0.23^{+0.06}_{-0.11}$&$-0.02^{+0.15}_{-0.13}$\\
Ursa Minor&$0.56^{+0.05}_{-0.05}$&$0.03^{+0.41}_{-0.18}$&$-0.25^{+0.06}_{-0.12}$&$-0.03^{+0.08}_{-0.11}$\\
Reticulum II&$0.59^{+0.02}_{-0.03}$&$0.08^{+0.29}_{-0.23}$&$-0.26^{+0.07}_{-0.10}$&$-0.04^{+0.08}_{-0.12}$\\
Ursa Major II&$0.63^{+0.05}_{-0.05}$&$0.06^{+0.32}_{-0.22}$&$-0.26^{+0.06}_{-0.09}$&$-0.04^{+0.07}_{-0.13}$\\
Hercules&$0.68^{+0.08}_{-0.08}$&$0.08^{+0.35}_{-0.23}$&$-0.24^{+0.06}_{-0.09}$&$-0.05^{+0.07}_{-0.10}$\\
Ursa Major I&$0.80^{+0.04}_{-0.04}$&$0.02^{+0.39}_{-0.19}$&$-0.22^{+0.08}_{-0.08}$&$-0.06^{+0.06}_{-0.10}$\\
\hline
\end{tabular}

\label{TableTriax}
\end{table*}

\section{Discussion and Conclusions}

Flattening is a crucial attribute of a dwarf spheroidal galaxy. Both
the dark halo and the stellar distribution can be flattened. The ultrafaint dSphs have many fewer baryons than the classical dwarfs so it is anticipated that feedback effects have a weaker effect on the shape of the dark matter distribution in the ultrafaints. Therefore, for the ultrafaints, a flattened stellar distribution probably corresponds to a flattened dark matter distribution. Of these ultrafaints, Reticulum~II is an interesting object as it is particularly nearby and also one of
the most highly flattened of all the ultrafaints, at least as judged
by the stellar light. On these grounds, we might well expect that
flattening may provide an explanation as to why a gamma-ray signal may have been seen towards Reticulum~II as opposed to other ultrafaints.

We have explored the impact of flattening on the J- and D-factors, which control the expected dark matter annihilation and decay signals from the dSphs. The effects of flattening on these factors can be decomposed into two separate corrections: the geometric and the kinematic factors. The first of these corresponds to the increase (decrease) in dark-matter density produced by squeezing (stretching) the models. The latter corresponds to how the observed velocity dispersion relates to the total velocity dispersion or the enclosed dark matter mass. When viewing oblate (prolate) models face-on, these two factors act together to increase (decrease) the J-factor over a spherical analysis, whereas, when viewing these models edge-on, the two factors compete and result in a decrease (increase) in the J-factor over a spherical analysis.

We have used Made-to-Measure techniques ~\citep{Sy96,De09} to
build numerical equilibrium models of Reticulum~II. These reproduce the flattened
shape, the major-axis length and the line of sight velocity dispersion of Reticulum~II. For the models with a prolate dark matter halo with ellipticity $\sim
0.6$ viewed edge-on, flattening could cause an additional
amplification of $\sim 2-2.5$ for Reticulum~II over that expected
for spherical dark halos. This factor could be still larger if the
stellar profile falls off more slowly than a Plummer law (which could
increase the kinematic factor). It could also be larger if the dark halo of Reticulum~II is triaxial (as anticipated from dark-matter-only simulations) and hence more flattened along the line-of-sight. However, this scenario is disfavored by dark-matter simulations with and without baryons that produce subhalos which preferentially point towards the center of their host halo and so we might anticipate dSphs to be elongated along the line-of-sight.

We corroborated the results of the Made-to-Measure simulations with a simpler virial method that allows for more rapid calculation of the correction factors for general geometries. A simple fitting relation has been provided for rapid estimation of the correction factors for the oblate and prolate cases. Additionally, we have inspected two cases where some analytic progress can be made in the computation of the J-factors. This has allowed us to characterize how the correction factors change as a function of the stellar and dark-matter distributions. We found that the correction factors for the Made-to-Measure models agree well with the trends seen in the analytic models.

We used our models to estimate the J-factors for the dSphs under the assumption that the figures are aligned with the line-of-sight and are either oblate or prolate. The ranking of the J-factors of the dSphs is slightly altered when accounting for flattening under the assumption that all the dSphs are either prolate or oblate. Typical correction factors for a dSph with ellipticity $0.4$ are $0.75$ in the oblate case and $1.6$ in the prolate case. We also demonstrated that the corrections to the D-factors are much smaller than the scatter in the spherical D-factor from the other observables. For instance, a dSph with ellipticity $0.4$ has a D-factor correction factor of $0.97$ in the oblate case and $1.1$ in the prolate case. Therefore, we concluded that flattening is unimportant for D-factor computation.

We concluded our discussion of the effects of flattening by computing correction factors for triaxial figures. The findings from the axisymmetric cases were found to simply extend when considering triaxiality. The largest J-factor correction factor corresponds to viewing the figure along the minor axis, whilst the smallest corresponds to viewing the figure along the major axis. We found that for a Reticulum~II-like model the J-factor correction factor varies by a factor of $\sim 6-10$ as one changes the viewing angle. For a fixed observed ellipticity, the correction factor can vary by a factor of $\sim4$. We demonstrated that for the known dSphs the uncertainty in the correction factors due to unknown triaxiality increases with the observed ellipticity of the dSph and is typically a factor of two for $\epsilon\sim0.4$. If all dSphs have their major axes aligned with the line-of-sight (as suggested by some simulations), the correction factors decrease as a function of observed ellipticity and are typically a factor $1/2$ for $0.4\lesssim\epsilon\lesssim0.6$.

Deviations from sphericity in both the light profile and the dark
matter are important. This suggests fundamental limitations to the
spherical Jeans modeling which is common in the field (although see \citep{Hayashi2016} for J-factors computed using axisymmetric Jeans modeling). In
particular, increasingly sophisticated statistical
techniques~\citep{Mar15,Co15} will fail to include an inherent uncertainty if the assumption of a
spherical stellar density profiles in a spherical dark halo breaks
down. The uncertainties, which are different for different dSphs, must be accounted for joint analyses of multiple dSphs.

Spherical Jeans modeling is probably most useful for large
classical dwarf spheroidal galaxies that look nearly round (such as
Leo I or Fornax). It ignores important uncertainties for the ultrafaints, which is
unfortunate as these are the most promising targets of all for
indirect dark matter detection. We hope that the work presented here
-- a systematic foray into the domain of flattening -- is the
beginning of a systematic exploration of more general flattened and
triaxial dark halo shapes.

Finally, this study was partly inspired by the gamma-ray detection~\citep{2015PhRvL.115h1101G} toward the very flattened ultrafaint, Reticulum~II. Our work demonstrates that Reticulum~II could have a J-factor that is higher than spherical analyses suggest if it is a prolate figure. However, the correction for the prolate shape does not make Reticulum~II stand out as the dSph with the highest J-factor nor does a lack of signal from the other dSphs create any tension, irrespective of the shapes of the other dSphs. If, however, Reticulum~II is an oblate figure the J-factor is lower than that found through spherical analyses and lack of signal from the other dSphs may give rise to some tension if the other dSphs (such as Ursa Minor) are prolate. More generically we have demonstrated that unknown triaxiality produces uncertainty in the J-factor for Reticulum~II of a factor of $\sim2$.

In general, we have shown that the effect of flattening on expected dark matter annihilation fluxes cannot be ignored. Indeed, flattening can shift expected signals by amounts larger than error bars due to current velocity dispersion measurements. These currently unknown shifts change the ranking of dSph targets for gamma-ray experiments. However, if the orientations of the Milky Way dSphs can be determined, the results presented here can help pin down relative J-factors and allow tests of dark matter explanations of gamma-ray detections.

\begin{acknowledgments}
JLS acknowledges financial support from the Science and Technology Facilities Council (STFC) of the United Kingdom. Figure~\ref{fig:flattened} was produced using the Pynbody package \citep{Pynbody}.
\end{acknowledgments}
\bibliography{bibliography}
\bibliographystyle{apsrev}

\appendix
\section{Virial ratios for cusped models}\label{CuspAppendix}
In Section~\ref{AxiCusps} we presented the virial ratio $\langle\sigma^2_{xx}\rangle/\langle\sigma^2_{zz}\rangle$ for a flattened cusped model with density slope $\gamma_\star=3$. Here we provide formulae for other values of $\gamma_\star$. For $\gamma_\star=2$, the virial ratio is
\begin{equation}
{\langle\sigma^2_{xx}\rangle \over \langle\sigma^2_{zz}\rangle}=\frac{Q_\star T(\mathcal{Q})-\mathcal{Q}}{2[\mathcal{Q}-T(\mathcal{Q})]},
\end{equation}
where $T=\arctanh$ for $Q_\star<1$ and $T=\arctan$ for $Q_\star>1$, and $\mathcal{Q}$ is defined in equation~\eqref{curlyQ}.
For $\gamma_\star=4$, the virial ratio is
\begin{equation}
{\langle\sigma^2_{xx}\rangle \over \langle\sigma^2_{zz}\rangle}=
\frac{Q_\star [ ( \mathcal{Q}^2-1)T(\mathcal{Q}) +
    \mathcal{Q}]}{2[Q_\star T(\mathcal{Q}) -\mathcal{Q}]}
\end{equation}
where we have corrected a typographical error in eq.~(21) of \citet{Ag12}. For $\gamma_\star=5$, the virial ratio is simply
\begin{equation}
{\langle\sigma^2_{xx}\rangle \over \langle\sigma^2_{zz}\rangle}=Q_\star.
\end{equation}
A plot of the ratios as a function of flattening are given as Figure 1
in \citep{Ag12}

\section{Velocity dispersions for flat rotation curve models}

In Section~\ref{CoredModels}, we presented formulae for the velocity
dispersions of a cored stellar profile with outer density slope
$\beta_\star=5$ embedded in a cored dark matter density profile. Here
we provide equivalent formulae for the cases $\beta_\star=4$ and
$\beta_\star=6$.

Let us recall the definitions
\begin{equation}
\Delta_1^2 = q_\phi^2 -q_\star^2,\qquad\qquad \Delta_2^2 = \Rd^2-\Rc^2,
\end{equation}
and let us introduce the function
\begin{equation}
F ={1\over \Delta_1 \Delta_2}\Bigl[
\arctan \left({\Rd \Delta_1\over q_\star \Delta_2}\right)-
\arctan \left({\Rc \Delta_1 \over q_\phi \Delta_2} \right)\Bigr].
\end{equation}
Notice that if $\Rc>\Rd$ or $q_\star>q$, this function remains
well-defined on using the identity (for $S<0$)
\[
{1\over \sqrt{S}}\arctan\sqrt{S} \equiv {1\over \sqrt{-S}}\arctanh\sqrt{-S}.
\]

For the case $\beta_\star=4$,
\begin{eqnarray}
\langle\sigma^2_{RR}\rangle &=&  {v_0^2 q_\phi\Rc \over
  \Delta_1^2\Delta_2^2} \Bigl[ (q_\phi^2\Rd^2 + q_\star^2(\Rc^2\!-\!2\Rd^2))F
  \\&-& q_\phi\Rc + q_\star\Rd\Bigr],\nonumber\\
\langle\sigma^2_{zz}\rangle &=&
{v_0^2 q_\star^2 \Rc\over \Delta_1^2} \Bigl[ q_\phi F - {q_\star \over
q_\star\Rc + q_\phi \Rd}\Bigr].
\end{eqnarray}

If $\Rc =\Rd$, then the velocity dispersions are a lot simpler. A
careful Taylor expansion gives
\begin{eqnarray}
\langle\sigma^2_{RR}\rangle &=& {2q_\phi(2q_\star+q_\phi)  v_0^2 \over
  3(q_\star + q_\phi)^2},\nonumber\\
\langle\sigma^2_{zz}\rangle &=& {q_\star^2v_0^2\over (q_\star + q_\phi)^2}.
\end{eqnarray}
so that when $q_\star = q_\phi$, we obtain
\begin{equation}
\langle\sigma^2_{RR}\rangle = {v_0^2 \over 2},\qquad\qquad
\langle\sigma^2_{zz}\rangle = {v_0^2\over 4}.
\end{equation}

Finally, we give the results for $\beta_\star =6$,
\begin{eqnarray}
\langle\sigma^2_{RR}\rangle &=&  {v_0^2 q_\phi\Rc^2 \over
 \Delta_1^2\Delta_2^4} \Bigl[
{ \Rd(\Rc^2+2\Rd^2)\Delta_1^2  -q_\star q_\phi\Rc\Delta_2^2 \over
    q_\star\Rc + q_\phi\Rd}\nonumber\\
&-&\Rc(3q_\phi^2\Rd^2+q_\star^2(\Rc^2-4\Rd^2))F
\Bigr], \\
\langle\sigma^2_{zz}\rangle &=&
{v_0^2 q_\star^2 \Rc^2 \over \Delta_1^2\Delta_2^2}
\Bigl[ -q_\phi \Rc F
+ {q_\star q_\phi\Rc\Rd + q_\phi^2\Rd^2 - q_\star^2\Delta_2^2 \over
(q_\star\Rc + q_\phi \Rd)^2}\Bigr].\nonumber
\end{eqnarray}
If $\Rc = \Rd$, then
\begin{eqnarray}
\langle\sigma^2_{RR}\rangle &=& 2v_0^2 {q_\phi(8q_\star^2 +
  9q_\star q_\phi + 3q_\phi^2)\over 15 (q+ q_\phi)^3},
\nonumber \\
\langle\sigma^2_{zz}\rangle &=& v_0^2 {q_\star^2(3q_\star+q_\phi)\over
  3(q_\star+q_\phi)^3}.
\end{eqnarray}
and if additionally $q_\star = q_\phi$, we recover
\begin{equation}
\langle\sigma^2_{RR}\rangle = {v_0^2 \over 3},\qquad\qquad
\langle\sigma^2_{zz}\rangle = {v_0^2\over 6}.
\end{equation}

\end{document}